\documentclass[twocolumn,superscriptaddress,nofootinbib,floatfix,aps,prb,citeautoscript,longbibliography]{revtex4-2}
\usepackage[utf8]{inputenc}
\usepackage[T1]{fontenc}
\usepackage{lineno}

\usepackage{hyperref}
\usepackage{graphicx}
\usepackage{color}
\usepackage{array}
\usepackage{dcolumn}
\usepackage{bm}
\usepackage{multirow}
\usepackage[version=4]{mhchem}
\usepackage{cleveref}
\usepackage{amsfonts}

\usepackage{orcidlink}

\begin{document}
\newcommand{\bochum}{Research Center Future Energy Materials and Systems of the University Alliance Ruhr and Interdisciplinary Centre for Advanced Materials Simulation, Ruhr University Bochum, Universitätsstraße 150, D-44801 Bochum, Germany}

\title{A generative material transformer using Wyckoff representation}

\author{Pierre-Paul De Breuck \orcidlink{0000-0002-3173-2058}}
\affiliation{\bochum}
\author{Hashim A. Piracha \orcidlink{0000-0002-1730-7995}}
\affiliation{\bochum}
\author{Gian-Marco Rignanese \orcidlink{0000-0002-1422-1205}}
\affiliation{UCLouvain, Institute of Condensed Matter and Nanosciences (IMCN), Chemin des Étoiles 8, Louvain-la-Neuve 1348, Belgium}
\author{Miguel A. L. Marques \orcidlink{0000-0003-0170-8222}}
\affiliation{\bochum}

\date{\today}

\begin{abstract}
    Materials play a critical role in various technological applications. Identifying and enumerating stable compounds—those near the convex hull—is therefore essential. Despite recent progress, generative models either have a relatively low rate of stable compounds, are computationally expensive, or lack symmetry. In this work we present Matra-Genoa, an autoregressive transformer model built on invertible tokenized representations of symmetrized crystals, including free coordinates. This approach enables sampling from a hybrid action space. The model is trained across the periodic table and space groups and can be conditioned on specific properties. We demonstrate its ability to generate stable, novel, and unique crystal structures by conditioning on the distance to the convex hull. Resulting structures are 8 times more likely to be stable than baselines using PyXtal with charge compensation, while maintaining high computational efficiency. We also release a dataset of 3 million unique crystals generated by our method, including 4,000 compounds verified by density-functional theory to be within 0.001~eV/atom of the convex hull.
\end{abstract}

\maketitle

\section{Introduction}

Inorganic solid materials are fundamental to numerous current and emerging technologies including energy storage, semiconductors, high-entropy alloys, catalysis, and \ce{CO2} capture. While many structures have already been enumerated thanks to experimental studies, the development of density functional theory (DFT)~\cite{hohenbergInhomogeneousElectronGas1964a, kohnSelfConsistentEquationsIncluding1965}, high-throughput methods, and open materials databases~\cite{curtaroloHighthroughputHighwayComputational2013a}, the discovery of novel materials still remains a central challenge in materials science. This pursuit is driven by the potential to replace existing materials with non-toxic, cost-effective, or lighter alternatives, or to uncover entirely new properties. The search space for novel materials is vast due to the immense number of possible combinations; however, the subset of materials that are thermodynamically stable ---defined as being near the convex hull of stability--- is significantly smaller, though still substantial.

To come up with novel candidates, the community has relied on various approaches including random sampling~\cite{pickardInitioRandomStructure2011}, evolutionary algorithms~\cite{oganovCrystalStructurePrediction2006,wangCrystalStructurePrediction2010}) and prototype substitution based on element similarity~\cite{hautierDataMinedIonic2011b} among others. These approaches are however computationally intensive and tend to yield low success rates. More recently, generative machine learning (ML) techniques have emerged as powerful tools for exploring novel stable structures. These models aim to learn the subspace of stable arrangements, but challenges remain due to the complexity of the three-dimensional (3D) structure of materials, their periodicity, and the vast scope of the periodic table.

Initial machine learning approaches use 3D voxel representations~\cite{hoffmannDataDrivenApproachEncoding2019a,nohInverseDesignSolidState2019a,court3DInorganicCrystal2020b,longConstrainedCrystalsDeep2021b}, but these methods face limitations, including a lack of rotational invariance and low success rates. Alternative methods based on Euclidean coordinates, often using GANs, have also been explored (e.g., CrystalGAN~\cite{nouiraCrystalGANLearningDiscover2019a}, CubicGAN~\cite{zhaoHighThroughputDiscoveryNovel2021a}, CCDCGAN~\cite{longConstrainedCrystalsDeep2021b}). However, these approaches are not invariant to Euclidean transformations and rely on the inherently unstable GAN framework. Approaches using the reciprocal space have also been used (e.g., FTCP~\cite{renInvertibleCrystallographicRepresentation2022b}). More recent advancements use diffusion models (e.g., DiffCSP~\cite{jiaoCrystalStructurePrediction2024}, CDVAE~\cite{xieCrystalDiffusionVariational2022a}, MatterGen~\cite{zeniGenerativeModelInorganic2025}), showing significant promise but lack explicit symmetry incorporation and are comparatively slower. Other strategies leverage large language models (LLMs) to learn crystallographic information file (CIF) representations, achieving success but requiring larger models with up to 200 million parameters (e.g., CrystalLLM~\cite{antunesCrystalStructureGeneration2024}). Variational autoencoders (VAEs) that incorporate Wyckoff position information have also been proposed, but they lack coordinate awareness (e.g., WyCryst~\cite{zhuWyCrystWyckoffInorganic2024}). Additionally, many of these methods do not embed stability considerations as an inductive bias, limiting their practical applicability. \\ \vspace{-0.3cm}\\
\indent In this work, we introduce Matra-Genoa, an efficient, scalable, generative materials transformer designed for the inverse design of inorganic crystal structures. Matra-Genoa is based on a sequenced Wyckoff representation of crystal structures, explicitly including atomic coordinates, making it, to our knowledge, the first generative model combining discrete Wyckoff positions and continuous coordinates. The framework also enables conditional generation based on target properties; in this study, we condition on energy above the convex hull to prioritize stable compounds. We first detail the representation, sampling, and conditioning mechanisms and then benchmark Matra-Genoa's capacity to discover stable, unique, and novel (S.U.N.) compounds. Finally, we demonstrate the model’s scalability by generating 3 million structures and assess their stability distribution.

\section{Results}
\subsection{Representation}

In materials science one can typically describe a material $m$ by a set of atoms and their Cartesian coordinates. This results in a high dimensional space $x$, encompassing an infinite amount of combinations. However, only a fraction of these combinations correspond to local energy minima, and an even smaller subset is thermodynamically stable. The current objective is to model $p(x)$ and more specifically its distribution modes. This probability distribution represents the likelihood of elements being compatible with binding at energetically favorable coordinates.

\begin{figure*}[ht!]
    \centering
    \includegraphics[width=0.7\linewidth]{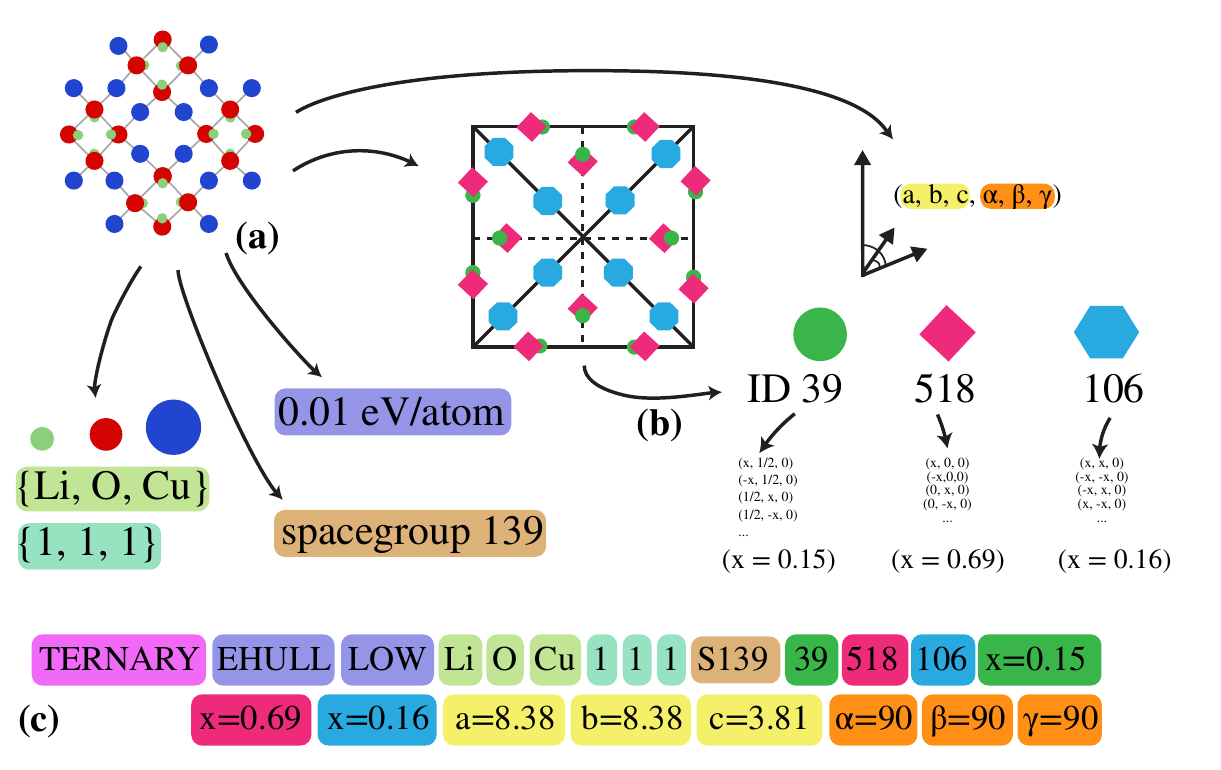}
    \caption{Schematic overview of the invertible sequenced representation. (a)~The structure is first decomposed into composition, stability, structure and lattice. (b)~The structure is then further decomposed into a set of Wyckoff positions, uniquely identified by a set of Wyckoff identifiers. Optional free parameters are also included to make the representation coordinate-aware. (c) All previous information is gathered into a tokenenized and invertible sequence. The color of the tokens represent the type or the Wyckoff position for ease of visualization.}
    \label{fig:matra_encoding}
\end{figure*}

To facilitate learning, we choose a coarse-grained representation of crystal structures as a first step, simplifying the system and making it easier to capture the distribution. We take inspiration from cross-disciplinary approaches in Natural Language Processing (NLP) and crystallography. Let a material $m$ be represented by $S_N=\{w_i \}_{i=1}^N$, a sequence of $N$ input tokens, that can be of different type, specifically specifying the composition, stoichiometry, structure and stability, see \cref{fig:matra_encoding}(a).

A key aspect is to develop an \textit{invertible} representation for the structure.
Taking inspiration from crystallography, we introduce an approach by considering Wyckoff positions, including free parameters. Without loss of information, any crystal structure can be described through (i)~the spacegroup, (ii)~a set of Wyckoff positions and corresponding chemical elements, (iii)~free parameters, if required, of the Wyckoff positions, and finally (iv)~the dimensions of the unit cell ($a,b,c,\alpha,\beta,\gamma$). A Wyckoff position reduces a set of equivalent points (orbit) into a single point, by mapping the equivalent sites under the symmetry transformations of the given space group.

Across the 230 three-dimensional spacegroups, there are 1731 different Wyckoff positions. Intrinsically, each position describes an orbit, as documented in the International Tables for Crystallography~\cite{chantlerInternationalTablesCrystallography2024}. Certain positions reappear across multiple space groups. For instance, the orbit of points ${(0,y,0), (0,-y,0)}$ appears in spacegroups 10, 16, and 47. We decide to use a shared representation for common orbits, resulting in a total of 990 unique Wyckoff positions across all spacegroups. \Cref{fig:matra_encoding}(b) schematically represents this Wyckoff extraction. Ultimately, the orbit can be fully characterized by specifying the algebraic terms of the Wyckoff positions.

This search space is a hybrid discrete and continuous space, which is conjectured to be not uniformly distributed energetically wise. In other words, we expect that given the presence of certain elements on certain positions, some other elements will be more favourable to appear in some specific Wyckoff positions. Furthermore, the continuous part of the space, which considers the free parameters in the Wyckoff positions (i.e. the coordinates) and lattice parameters, is also expected to depend on sampled elements and positions. This non-uniformity is further discussed in \cref{sec:discussion} and \cref{fig:wyckoff_dist}, and forms a strong motivation to use a neural network to learn the underlying distribution.

Next, the token sequence $S_N$ is embedded as $E_N=\{x_i \}_{i=1}^N$, where $x_i\in R^d$ is the $d$-dimensional embedding vector of token $w_i$. Tokens are either encoded using learnable or Gaussian embeddings, depending on its type (e.g., elements, stoichiometry, Wyckoff letter, coordinate). Positional information is incorporated using sinusoidal functions. More information and architecture details are provided in \cref{sec:methods}.

\subsection{Training, sampling, and conditioning}

Given the previous representation $E_N$, one is now interested to learn how the tokens are distributed. We use a transformer architecture to learn how to sample the token $x_{n+1}$ based on all previous tokens $\{x_i \}_{i=1}^n$. For instance, the model completes the sequence "TERNARY EHULL LOW Ti" by various possible next tokens: O, Ca, Ni, $\ldots$, and continues with the spacegroup (e.g. $62, 225, \ldots$), Wyckoff positions (e.g. 4b, 4c, 8d) and free parameters if needed (e.g. $x=0.79$). As each token depends on the previous representations, the neural network, using multiple transformer blocks relying on attention mechanisms, learns a powerful representation of symmetrized crystals. Given the hybrid space, two prediction heads are used for discrete and continuous tokens. The network is trained using causal attention and a mixed cross-entropy means square error loss. Inference is done using a simple greedy approach, following a Boltzmann distribution:
\begin{equation}
    p(x_{n+1}) \propto \exp\left(\frac{f(\{x_i \}_{i=1}^n)}{T}\right)
\end{equation}
where $f(\{x_i \})$ are logits obtained from the transformer network.

In order to condition on thermodynamic stability, defined by the compound's energy distance to the convex hull, we inform the model by starting the sequences with a discrete stability token. The value of the token is set to either \textsc{low} or \textsc{high} depending on an arbitrarily fixed threshold, taken to be 0.075~eV/atom in this work. By starting the sequence with this information, the model knows \textit{a priori} what is expected. This enables it to go beyond a single distribution and learn to condition based on the energy above hull. This conditioning can, in principle, be extended to any property. However, in this work, we limit conditioning to stability, chemical space (defined by the elements), required symmetry (space group), and specific Wyckoff positions. This choice makes the proposed approach well-suited for investigating chemical spaces and exploring specific structural prototypes.

Two models are trained in this work, one on the Materials-Project (MP), named Matra-Genoa-MP, and one on the combined MP and Alexandria dataset, named Matra-Genoa-MPAS. More information about the data is provided in \cref{sec:data}. During training, 10\% of the sequences are left out to monitor various validation metrics. 

During training, a distributed representation is learned for the different elements, space groups, and Wyckoff positions. A t-distributed stochastic neighbor embedding (t-SNE) dimensionality reduction is applied for visualization, as shown in \cref{fig:elem_embedding} and~\cref{fig:spg_embedding} of the Supplemental Material. While the overall representation is broadly spread (indicating more complex patterns), meaningful clusters can still be observed, grouping elements into common categories. Although not investigated in this work, the learned embeddings could prove useful for supervised learning of materials properties.

\begin{figure}[ht!]
    \centering
    \includegraphics[width=\linewidth]{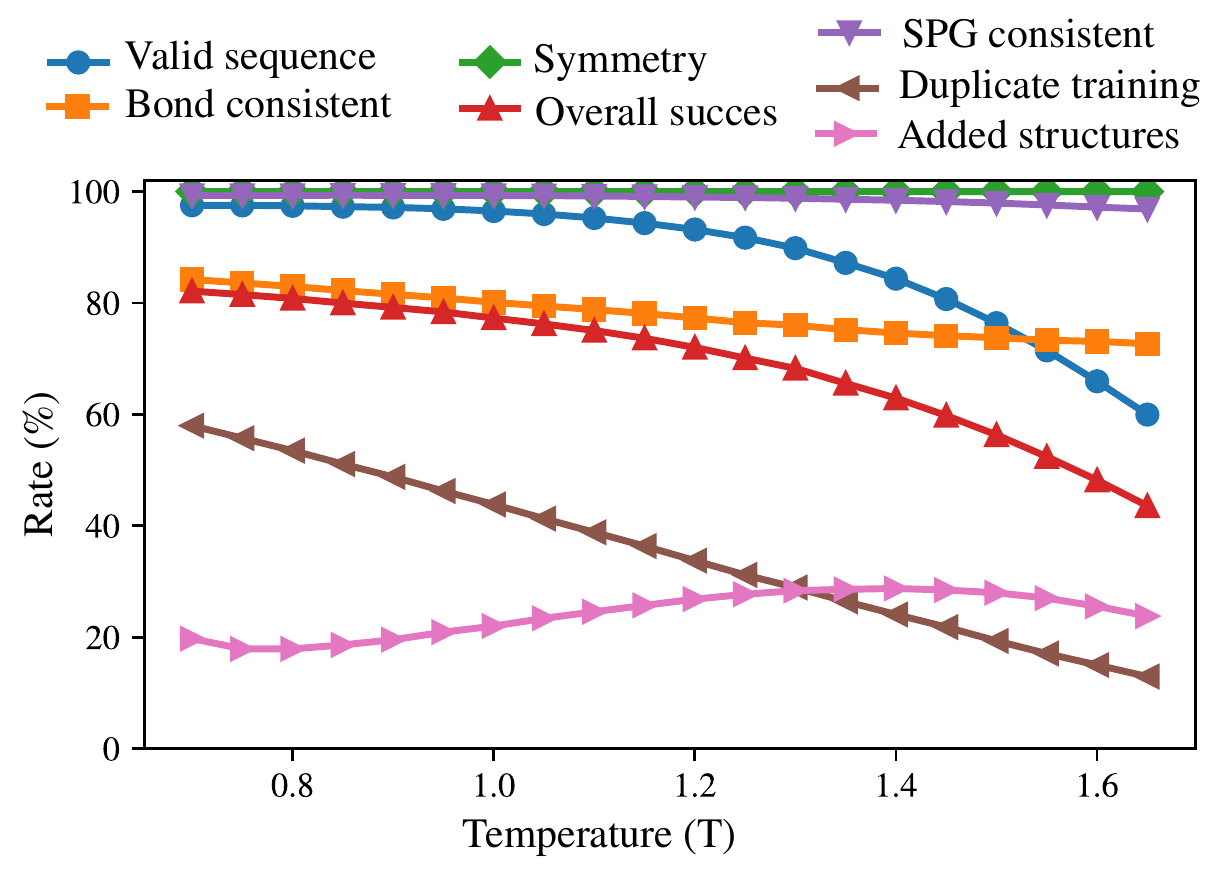}
    \caption{Success rates when generating novel compounds. The validity corresponds to sequences that can be converted to structures. Bond consistency refers to structures with no overlapping atoms ($<0.7$~\AA). SPG consistent structures have a spacegroup corresponding to the one explicitly sampled in the sequence.}
    \label{fig:success_rates}
\end{figure}

\Cref{fig:success_rates} presents various validity rates when sampling new compounds from the Matra-Genoa-MPAS model. Invalid sequences are defined as sequences that have either missing information (e.g. missing spacegroup, Wyckoff positions, free parameters or lattice) or incompatible Wyckoff positions, free-parameters or lattice parameters. Sequence validity ranges from 60\% at high sampling temperature up to 97\% at lower temperatures. For each structure, a site neighbour search is also performed, removing structures with colliding atoms (distance below $0.7$~\AA). We find that 28\% (high $T$) to 17\% (low $T$) of the structures have invalid bond lengths. All valid structures have symmetry (i.e. $\text{spacegroup} > 1$), and only a very small fraction (3\% at high $T$ to 0.7\% at low $T$) have a parsed spacegroup that differs from the one explicitly given in the sequence. This can happen when the sampled Wyckoff positions result in a higher symmetry group. Finally, we observe a considerable duplicate rate with the training set, that is around 57\% at $T=0.7$. This can be reduced by increasing the sampling temperature, with a duplicate training rate of 7.7\% at $T=1.65$, but at the cost of reducing the validity of the sequences to around 60\%. This suggests that one can create structures farther from the training set by increasing the sampling temperature. Overall, accounting for invalid sequences, invalid bonds, and duplicates, one can expect an added structure rate of 20\% to 30\% depending on the sampling temperature. Although this might seem low, this is more than sufficient, given how fast it is to generate and filter out invalid structures. In practice the final throughput of novel and unique structures far exceeds our requirements ($\sim$1000 unique and novel structures per minute). 

\subsection{Generating novel structures}

Matra-Genoa-MP was trained on the limited dataset comprising of the 
Materials Project, to investigate if it can successfully generate novel structures not seen during training. For this, we use the Wang-Botti-Marques (WBM) dataset (see \cref{sec:methods}) that has known stable structures not present in the Materials Projects. We evaluate the model's ability to recover these structures.

We first generate a series of compositions, to check if the model captures expected chemical tendencies. We observe that the vast majority contain O, Li or Mg, as it closely mimics the global tendencies in Materials Project (see Supplemental Materials for more details on the elemental and spacegroup distribution). From the 100,000 compositions, 41.6 \% are valid according to the rules of Semiconducting Materials from Analogy and Chemical Theory (SMACT). This is in line with the training set (the Materials Project has 47.2\% SMACT valid structures), without imposing any explicit composition rules.
Moreover when conditioning on elements less seen in the training (e.g. Be, Zn, In), we still observe sensible compositions with a reasonable SMACT validity (above or around 40\%). Without surprise, the model has learned a powerful compositional representation, without imposing any rules, and can easily go to compositions beyond the training data.

Going further to the structural part, we used the Al--Ca--Cu ternary space as an example. The Materials Project has only 3 structures in this space, respectively in spacegroups 139, 166 and 191. The WBM dataset has 15 structures on the hull for this space, in a variety of spacegroups (12, 63, 71, 57, 129, 166, 191, 194). Exhaustively searching all possible combinations of stoichiometries, spacegroups, and Wyckoff positions in this chemical system is computationally infeasible, with the combinatorial space exceeding $10^12$ potential structures.

\begin{figure}[h!]
    \centering
    \includegraphics[width=\linewidth]{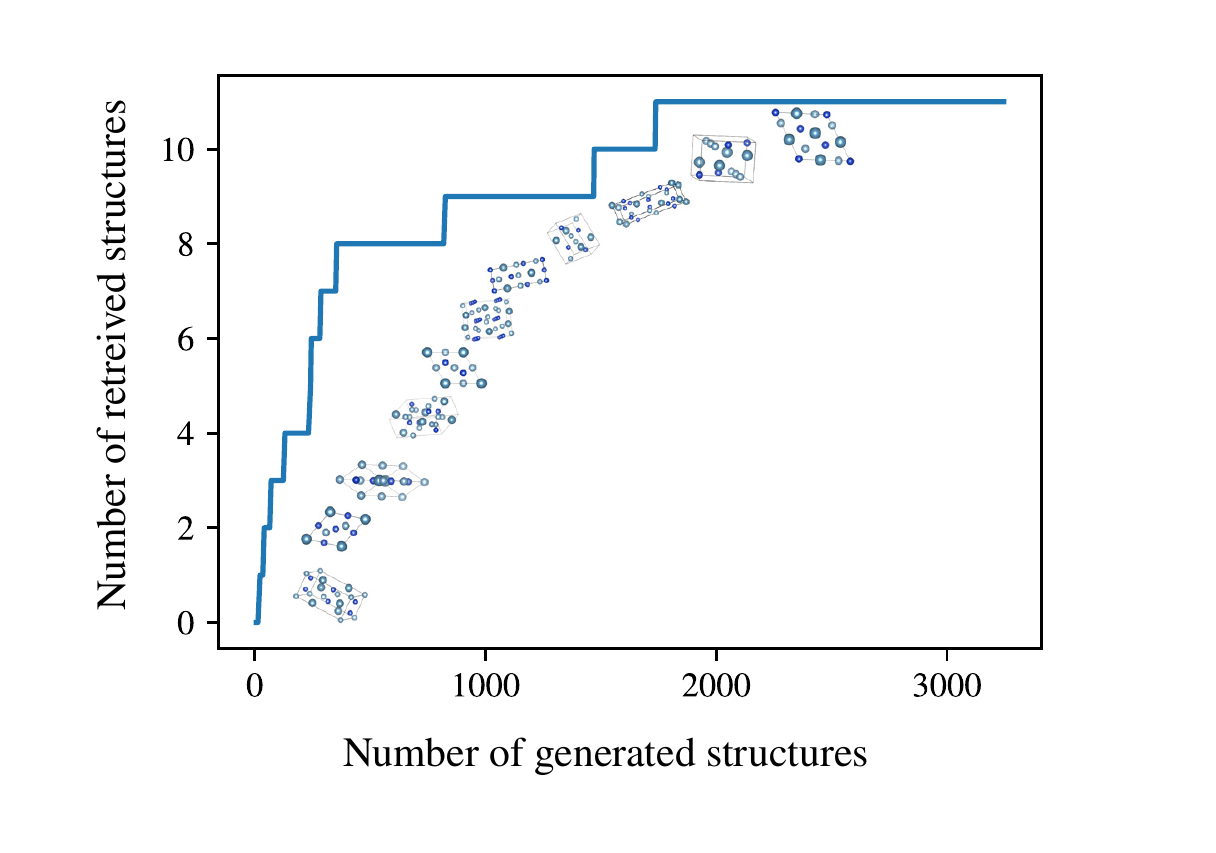}
    \caption{Searching for stable structures in the Al--Ca--Cu ternary space. Number of retrieved stable (on the convex hull) structures in WBM-test set with respect to the total number of generated structures.}
    \label{fig:WBM-test}
\end{figure}

We condition Matra-Genoa-MP on this chemical space, without telling the formula nor the spacegroup number. We are able to recover 11 stable structures (i.e. 73\%) within 2,000 generated structures. \Cref{fig:WBM-test} shows the number of retrieved structures versus generated candidates. This simple experiment shows that alongside being able to generate novel unseen compounds, it can also guess stable configurations with an reasonable efficiency, orders of magnitude faster than random sampling. The 4 structures that were not found all belong to spacegroups 12 or 194. 

\subsection{Going to millions of structures}

In order to go beyond and get a large set of novel realistic structures, we generate 3 million structures using the Matra-Genoa-MPAS model. In order to better understand the effect of sampling temperature, batches of 150,000 structures are generated at increasing sampling temperature, from $T=0.7$ to $T=1.65$ in steps of 0.05. We extend the generation pipeline by including a relaxation using computationally inexpensive universal machine learning interatomic potentials (uMLIPs). This enables one to have structures closer to their local energy minimum without requiring expensive computations, as well discarding nonphysical structures. The duplicate check is also extended by using a much larger set of known structures, namely the upstream candidates of the Alexandria dataset, comprising around 70 million structures optimized with the ORBITAL uMLIP, named ORB here. Stability is measured by estimating the energy above the hull using an ALIGNN model, relative to the Alexandria dataset. More details are found in \cref{sec:methods}.

\begin{table}[h]
\centering
\caption{Filtering results on the 3 million generated structures ($N_\text{batch} = 150,000$). Columns include sampling temperature $T$, successfully optimized structures $N_\text{opt}$, novel inserted structures $N_\text{novel}$, and stable structures meeting thresholds of 0.001, 0.050, and 0.100~eV/atom.}
\label{tab:3m}
\begin{ruledtabular}
\begin{tabular}{cccccc}
$T$ & $N_\text{opt}$ & $N_\text{novel}$ & $N_{0.001}$ & $N_{0.050}$ & $N_{0.100}$ \\
\hline
0.70 & 138425 & 57004 & 208 & 9441  & 24307 \\
0.75 & 137523 & 60952 & 219 & 9451  & 24482 \\
0.80 & 136892 & 64423 & 204 & 9265  & 24336 \\
0.85 & 136262 & 67096 & 246 & 9110  & 24214 \\
0.90 & 135678 & 69465 & 226 & 8717  & 23461 \\
0.95 & 135292 & 72438 & 233 & 8642  & 23378 \\
1.00 & 135054 & 74814 & 243 & 8215  & 22948 \\
1.05 & 134550 & 76499 & 218 & 8093  & 22448 \\
1.10 & 134215 & 78604 & 248 & 7848  & 22019 \\
1.15 & 133714 & 80541 & 213 & 7516  & 21523 \\
1.20 & 133544 & 82177 & 254 & 7406  & 21036 \\
1.25 & 133142 & 84319 & 243 & 7159  & 20593 \\
1.30 & 132838 & 85746 & 204 & 6831  & 19954 \\
1.35 & 132809 & 87650 & 232 & 6731  & 19724 \\
1.40 & 132246 & 89009 & 219 & 6500  & 19227 \\
1.45 & 131841 & 90082 & 200 & 6367  & 18773 \\
1.50 & 131634 & 91661 & 232 & 6071  & 18137 \\
1.55 & 131514 & 92441 & 195 & 5908  & 17797 \\
1.60 & 131326 & 93922 & 210 & 5641  & 17377 \\
1.65 & 131002 & 94764 & 198 & 5642  & 17063 \\
\end{tabular}
\end{ruledtabular}
\end{table}

\Cref{tab:3m} shows the amount of structures left after relaxation, duplication and stability filtering. $N_\text{opt}$ gives the number of successful relaxed structures. We observe that around 10\% fail due to mostly disconnected parts (distance between atoms larger than 1.5 times the sum of the covalent radii of the respective chemical elements). Next, $N_\text{novel}$ gives the number of unique and novel structures, with respect to the \textsc{ORB} dataset. Here the majority of compounds are lost (59\% at $T=0.7$ to 28\% at $T=1.65$), as \textsc{ORB} is a quite extensive set. As previously observed, increasing temperature considerably increases the novelty of the structures with less duplicates. Finally, $N_{0.001}$, $N_{0.050}$, $N_{0.100}$ represents the number of compounds respectively below 0.001, 0.050 and 0.100~eV/atom above the convex hull. It turns out that 42\% (at $T=0.7$) to 18\% (at $T=1.65$) of the inserted structures are below 0.001~eV/atom, which is a considerable amount. Overall, Matra-Genoa-MPAS has a S.U.N ratio of around 16\% at $T=0.7$. For comparison, MatterGen~\cite{zeniGenerativeModelInorganic2025} reports a S.U.N ratio of $\sim$45\%, CDVAE~\cite{xieCrystalDiffusionVariational2022a} of $\sim$18\%, and P-G-SchNet, G-SchNet~\cite{gebauerSymmetryadaptedGeneration3d2020} and FTCP~\cite{renInvertibleCrystallographicRepresentation2022b} are below 5\%, although these results are based on a much less stringent reference set (Alex-MP-ICSD) than ours.

\begin{figure}[ht!]
    \centering
    \includegraphics[width=\linewidth]{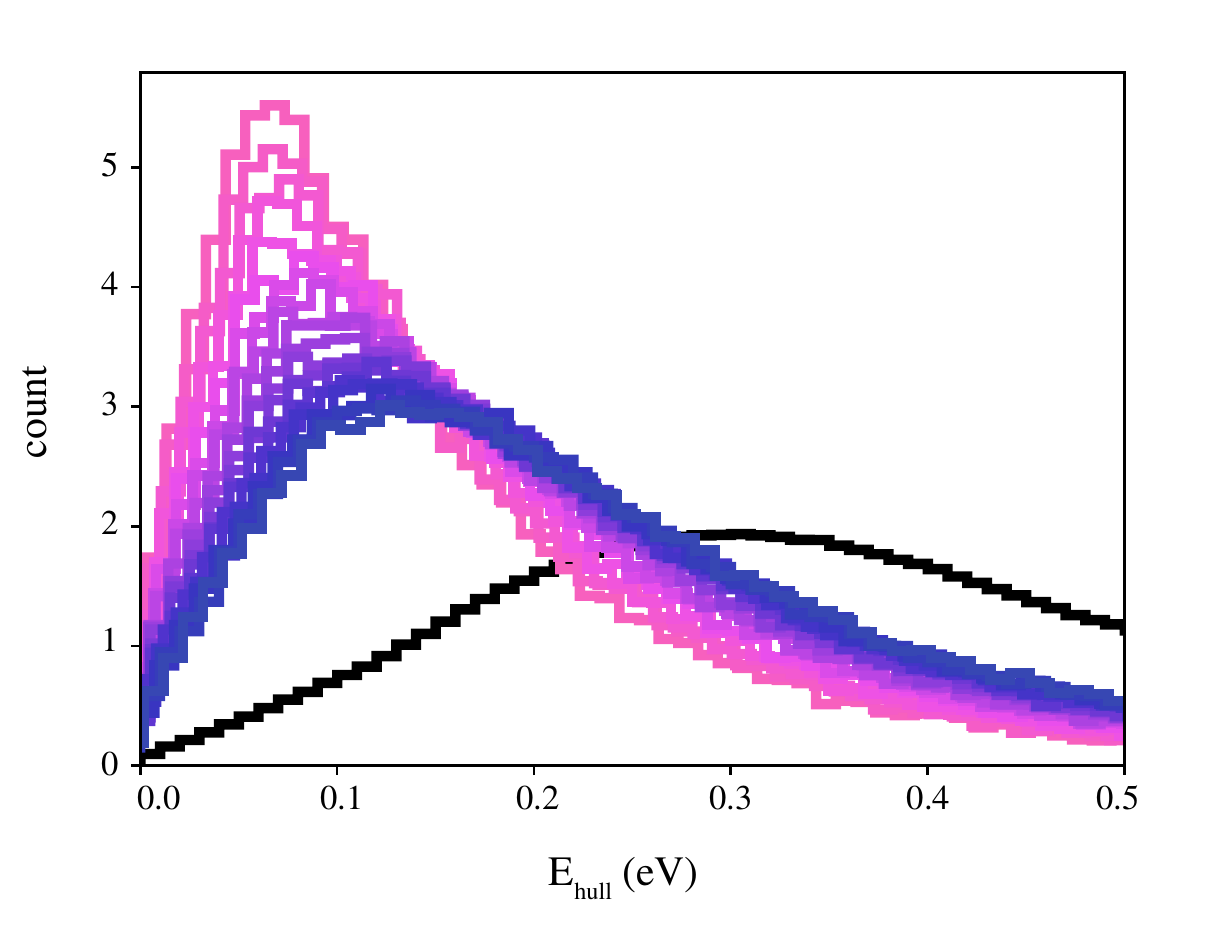}
    \caption{Distribution of distances to the convex hull as a function of the sampling temperature $T$ used to generate the structures. The values of $T$ change linearly from 0.70 (light pink curve) to 1.65 (dark blue). The black line represents results from random search with PyXtal~\cite{fredericksPyXtalPythonLibrary2021} imposing charge compensation. To ease the comparison, all curves are normalized to one.}
    \label{fig:ehull_dist}
\end{figure}

\Cref{fig:ehull_dist} illustrates the distribution of computed energies above the hull for the inserted structures across various sampling temperatures. For comparison, the distribution of structures generated by PyXtal~\cite{fredericksPyXtalPythonLibrary2021}, incorporating charge compensation, is shown in black. Notably, a pronounced non-uniform distribution favoring lower energy values is observed, with this trend becoming more pronounced at lower sampling temperatures. Compared to the baseline, a considerable improvement is found. This demonstrates the capability of Matra-Genoa to effectively generate structures with low-energy configurations.

From the 3 million generated structures, we select 15,0000 structures with the lowest estimated energy above the convex hull, and perform density functional theory (DFT) calculations for them (for computational details, see \cref{sec:methods}). Of the 13,249 converged calculations, 12,612 structures have distances to the hull below 0.050~eV/atom, and 4,094 are below 0.001~eV/atom.

\begin{figure}[]
    \centering
    \begin{tabular}{ccc}
        \includegraphics[height=.3\columnwidth]{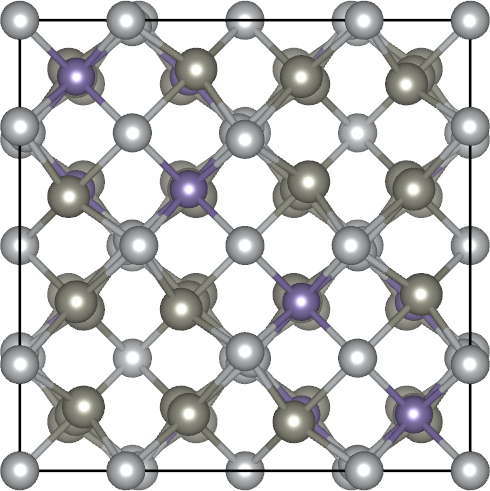} &
        \includegraphics[height=.3\columnwidth]{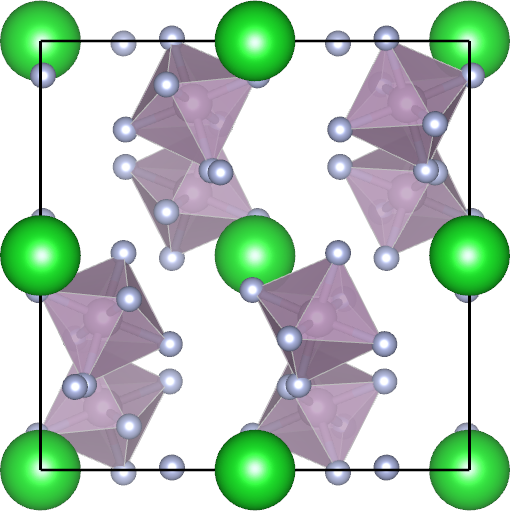} &
        \includegraphics[height=.3\columnwidth]{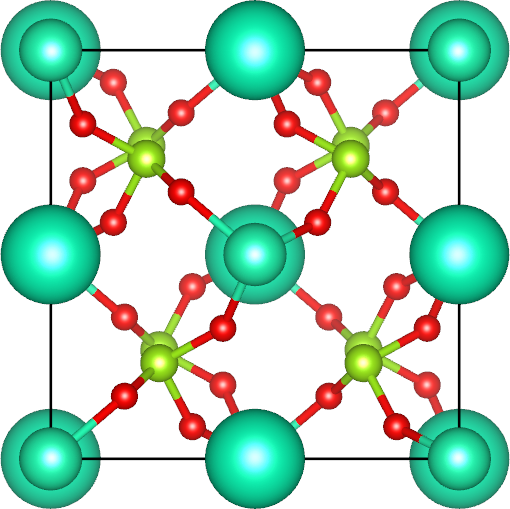} \\
        \ce{Zn6Ni7Ge2} & \ce{BaP2F12} & \ce{CsLu(SeO3)2} \\
        (spg. 227) & (spg. 205) &  (spg. 205) \\[3mm]
        \includegraphics[height=.3\columnwidth]{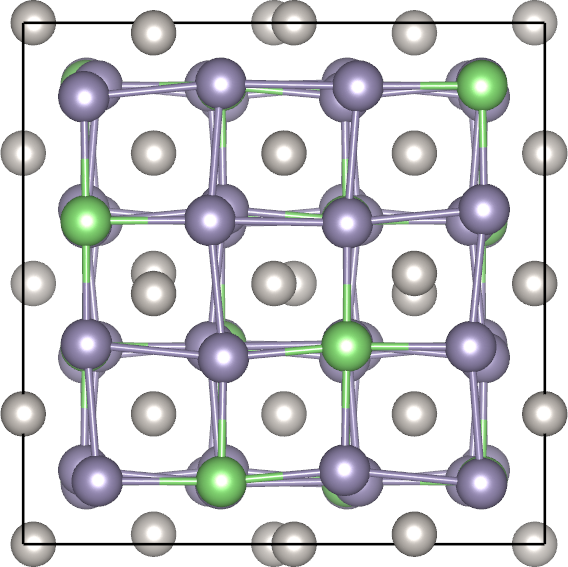} &
        \includegraphics[height=.3\columnwidth]{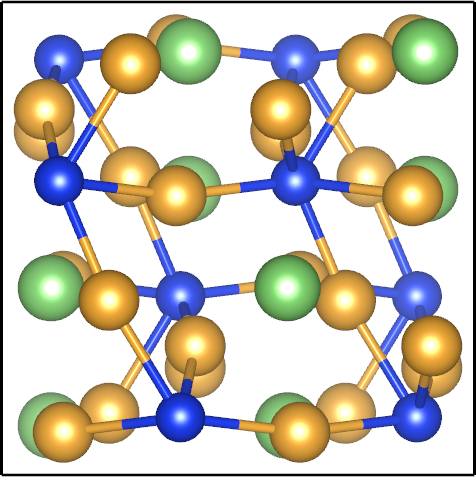} &
        \includegraphics[height=.3\columnwidth]{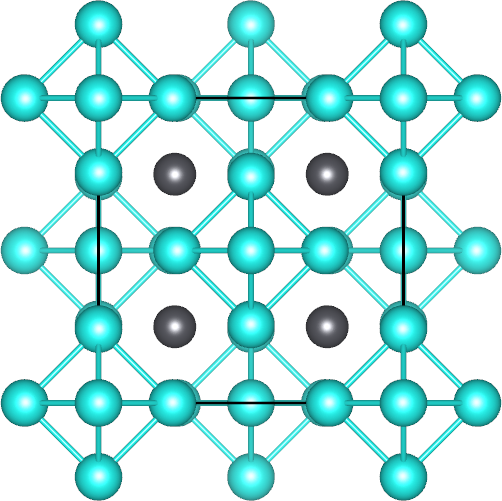} \\
        \ce{LiSn3Pt2} & \ce{LiSiAu3} & \ce{Yb6Pb2N} \\
        (spg. 206) & (spg. 205) & (spg. 225) \\[3mm]
        \includegraphics[height=.3\columnwidth]{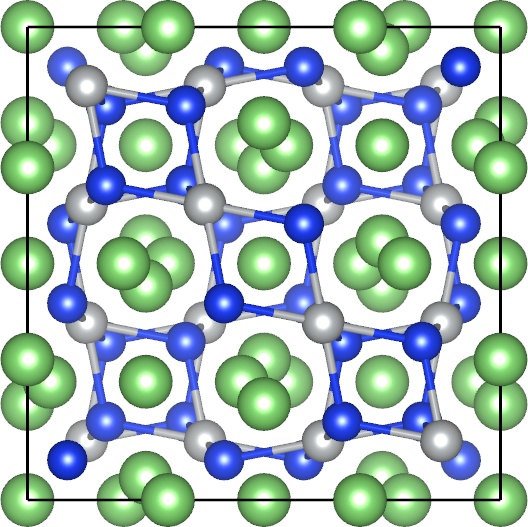} &
        \includegraphics[height=.3\columnwidth]{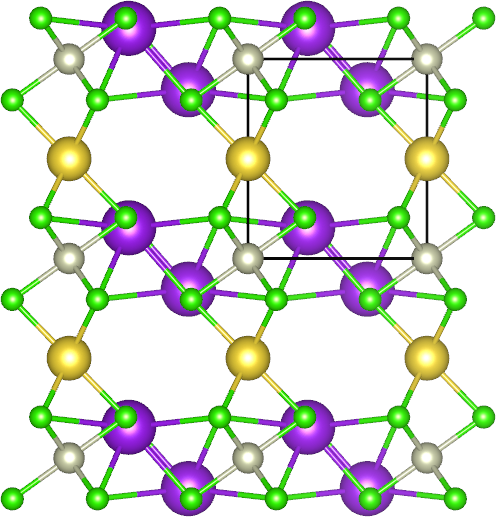} &
        \includegraphics[height=.3\columnwidth]{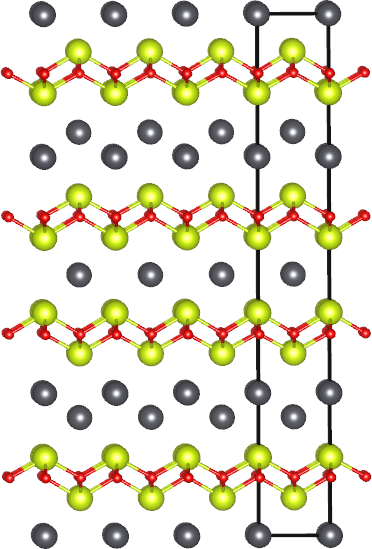} \\
        \ce{Li3Si2Ni} & \ce{K2NaRhCl6} & \ce{Ce4Pb3O4} \\
        (spg. 227)  & (spg. 164) & (spg. 139)
    \end{tabular}
    \caption{Example crystal structures in their conventional cells. Images produced in \textsc{VESTA}~\cite{mommaVESTA3Threedimensional2011}, with the default colors for the atoms.}
    \label{fig:crystals}
\end{figure}

\Cref{fig:crystals} shows example crystal structures from the latter group in conventional cell settings. We selected highly symmetric compounds that were not present on the convex hull of Alexandria. A variety of structures are observed, some of which display novel patterns not found in the training set. We discuss the 9 structures in more detail below.

The cubic phase of \ce{Zn6Ni7Ge2} is a crystalline stoichiometric coloring of the face-centered-cubic lattice, although we can expect that this chemical composition will be alloyed to some extent in experiment. There are only 35 other compounds with anonymous composition \ce{A2B6C7} in the convex hull of Alexandria, but none of these has cubic symmetry. The compound \ce{BaP2F12} crystallizes in a cubic phase with 60 atoms (4 formula units) in the primitive cell. It exhibits isolated \ce{PF6} octahedra embedded in a body-centered-cubic Ba framework. There are two compounds with related stoichiometries in the convex hull of Alexandria, specifically \ce{BaAs2F12}, \ce{SrAs2F12}, but in a structure with spacegroup 221 exhibiting different arrangement of the alkali earth metals. 

The generated quaternary phase of \ce{CsLu(SeO3)2} is obtained by substituting Y with Lu in the yttrium selenite \ce{CsY(SeO3)2}~\cite{bangVariableFrameworkStructures2014}. In this structure, Lu and Ce form a rocksalt sub-lattice, with \ce{SeO3} units located at the center of the cubes. For \ce{LiSn3Pt2}, we observe a simple cubic framework of Li and Sn, which may be disordered due to the similar ionic radii of Li and Sn in six-fold coordination, with some cubes containing Pt. This is the only compound in this prototype and chemical composition that appears on the convex hull of Alexandria.

The compound \ce{LiSiAu3} consists of vertex-sharing, distorted triangular prisms of Au with Si at the center, and Li intercalated in the empty spaces. This is the only compound in Alexandria with this prototype, and it is 0.070~eV/atom lower in energy than the lowest polymorph present in our Alexandria database. In \ce{Yb6Pb2N}, the N atoms form a face-centered cubic sublattice, with Yb atoms arranged in isolated octahedra around the cation, and with Pb exhibiting 12-fold coordination. The structure is reminiscent of an inverted perovskite. Although the inverted perovskite \ce{PbNYb3} is not found in Alexandria, we do observe similar structures with other rare earth elements, such as \ce{PbNCe3} and \ce{PbNNd3}, suggesting that the Yb inverted perovskite may also be stable or close to stability.

\ce{Li3Si2Ni} has 24 atoms in the primitive unit cell and crystallizes in the cubic space group number 227. It can be envisioned as a Si–Ni framework with Li intercalated within it. This is the only stable structure in Alexandria with this prototype and chemical composition. The trigonal structure of \ce{K2NaRhCl6} features layers of \ce{K2RhCl6} intercalated with layers of Na. This is the only stable quaternary structure in this prototype found in Alexandria. Interestingly, we also identify stable (or nearly stable) compounds with the compositions \ce{K3RhCl6} and \ce{Na3RhCl6}, although they have very different geometric arrangements. Additionally, we find the closely related composition \ce{K2NaIrCl6} near the hull of Alexandria, but in a distorted perovskite structure with \ce{NaCl6} and \ce{IrCl6} octahedra.

Finally, \ce{Ce4Pb3O4} has a unique structure, with \ce{CeO} layers alternating with flat and buckled Pb layers. This is the only tetragonal structure with the composition \ce{A3B4C4} found on the convex hull of Alexandria. Given the chemical similarity of the rare earth elements, it is likely that other elements can replace Ce in this structure to form stable compounds. The variety and novelty observed in these examples demonstrate that our pipeline, utilizing Matra-Genoa, is effective in identifying new compounds that are on or near the convex hull.

\section{Discussion}
\label{sec:discussion}

\begin{figure*}[ht!]
    \centering
    \includegraphics[width=\linewidth]{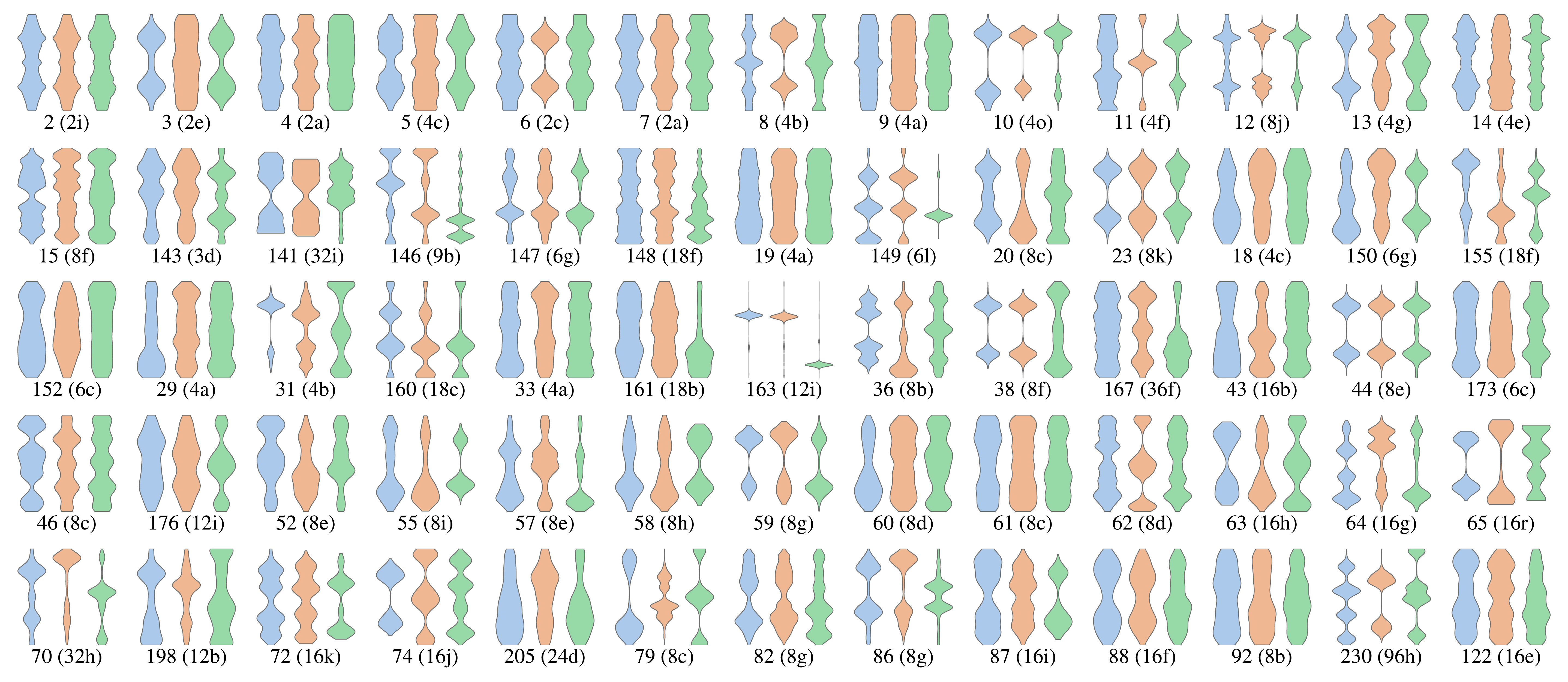}
    \caption{Distribution of the $x$, $y$, and $z$ coordinates for general Wyckoff positions in each space group with at least 1,000 occurrences, across all structures in Alexandria below 0.250~eV/atom and the Materials Project. Although all positions have three free parameters, structures show a preference for certain positions. The $x$-coordinate is shown in blue, $y$ in orange, and $z$ in green.}
    \label{fig:wyckoff_dist}
\end{figure*}
We introduced a novel generative machine learning framework for the inverse design of inorganic crystal structures through sequential sampling of Wyckoff coordinates. Wyckoff positions, rooted in crystallographic symmetry principles, provide an intuitive and robust framework for this work. They offer a natural representation of crystal formation and simplify the distribution of energetically favorable arrangements. \Cref{fig:wyckoff_dist} illustrates this idea by representing the $(x, y, z)$ coordinates of the unique general Wyckoff positions for each space group (i.e., positions invariant only under the identity operation). The coordinates show a non-uniform distribution in most cases, with more pronounced patterns in certain space groups (e.g., 10, 12, 146, etc.). Furthermore, one can expect that these distributions become even more localized when further conditioned on specific chemical subsystems due to constraints such as bond lengths and coordination numbers.

When trained on over 2 million structures, a powerful representation is learned and novel candidates can be generated by greedy sampling. The model was shown to successfully retrieve known stable structures within a ternary space. Furthermore, 3 million structures were generated under varying sampling temperatures. A trade-off between novelty and stability was observed: lower temperatures tended to reproduce the training set with more stable structures, whereas higher temperatures yielded greater structural diversity at the expense of stability, resulting in broader energy distributions.

The proposed approach offers several advantages. It inherently enforces symmetry constraints, is computationally efficient (training and sampling rely on a restricted shared vocabulary), is easy to understand and fully incorporates coordinates through a hybrid action space. Nevertheless, there is room for improvement. For instance, simple greedy sampling may not be the optimal strategy for conditioned sampling. Advanced techniques such as Monte Carlo Tree Search (MCTS), beam search, or GFlowNets~\cite{bengioFlowNetworkBased2021a} could enhance performance. Despite these limitations, the method’s rapid generation rate (up to 1,000 structures per minute) and the affordability of downstream machine learning screening make it highly practical. Future work could explore conditioning on additional material properties or adapting the framework for supervised learning.

In summary, by sampling from a coarse-grained representation of Wyckoff coordinates with a deep attention-based neural network, we demonstrated that it is possible to fully generate stable, crystal structures in an end-to-end manner (i.e. including coordinates). We hope this work inspires further advancements in inverse materials design and discovery.

\section{Methods}
\label{sec:methods}

\subsection{Data}
\label{sec:data}
\vspace{-5mm}
Four datasets are used for training and experiments in this work: MP, AS, WBM and ORB.

MP: The Materials Project as of 2023-02-07 (Creative Commons Attribution 4.0 International License), see Ref.~\onlinecite{jainMaterialsProjectMaterials2013}, is an open-access database containing DFT-relaxed crystal structures obtained mainly from experimentally-known crystals. The data is retrieved from matbench-discovery~\cite{MatbenchDiscovery}. Structures with more than 15 Wyckoff sites are removed. This amounts to a total of 115,663 structures.

AS: The Alexandria dataset (Creative Commons Attribution 4.0 International
License), see Ref.~\onlinecite{schmidtMachineLearningAssistedDeterminationGlobal2023}, is an open-access database containing DFT-relaxed crystal structures from
a variety of sources, including a large quantity of hypothetical crystal structures
generated by machine-learning methods. Data retrieved from Ref~\onlinecite{MatbenchDiscovery}. We filter out structures that are above 0.250~eV/atom of the convex hull and structures with more than 15 Wyckoff sites. This results in 2,553,057 samples.

In both previous datasets, we convert structures to their settings found in the International Tables for Crystallography~\cite{chantlerInternationalTablesCrystallography2024} and parse their Wyckoff positions using spglib~\cite{togoSpglibSoftwareLibrary2024}. Structures that are in spacegroup 1 are systematically discarded. Our final Matra-Genoa-MPAS model is therefore trained on 2,668,720 structures.

WBM: The Wang-Botti-Marques dataset, see Ref.~\onlinecite{wangPredictingStableCrystalline2021} contains 257,487 structures generated through single-element substitutions on Materials Project structures. Replacement is done based on chemical similarity matrix computed from the Inorganic Crystal Structure Database (ICSD)~\cite{bergerhoffInorganicCrystalStructure1983}.

We also use ORB dataset. Although not explicitly used for training, we use it to measure novelty (i.e. finding duplicates). This database currently comprises approximately 72 million entries, including 5.5 million compounds within 0.05~eV/atom of the convex hull and 14 million within 0.1~eV/atom. It incorporates compounds from the main Alexandria dataset, chemically substituted compounds~\cite{wangPredictingStableCrystalline2021}, and around 5 million randomly generated structures via PyXtal~\cite{fredericksPyXtalPythonLibrary2021}. These structrues are optimized with the ORBITAL uMLIP~\cite{neumannOrbFastScalable2024}

Finally, energy above hull estimates are computed using as reference the full Alexandria dataset, containing ~4.5M structures~\cite{Alexandria}.

\subsection{Encoding and transformer architecture}

The embedding $\mathbf{x}_i \in \mathbb{R}^d$ is either trained, or in case of continuous inputs, encoded as Gaussians:
\begin{equation*}
  \mathbf{x}_{\text{lin}} = \exp\left(- \frac{(r_{\text{lin}} - v)^2}{2\sigma^2}\right)  
\end{equation*}
\begin{equation*}
   \mathbf{x}_{\text{log}} = \exp\left(- \frac{(r_{\text{log}} - |\log(|v| + 10^{-20})|)^2}{2\sigma^2}\right) 
\end{equation*}
where $r_{\text{lin}}$ and $r_{\text{log}}$ are evenly spaced vectors in $\mathbb{R}^{d/2}$ and $\sigma$ a hyperparameter. The final embedding is:
\begin{equation*}
\mathbf{x}_i = \text{Concat}(\mathbf{x}_{\text{lin}}, \mathbf{x}_{\text{log}}).
\end{equation*}

Positional encodings $\mathbf{E}_{\text{pos}}$ are added to the embeddings to incorporate positional information, resulting in:
\begin{equation*}
\mathbf{z}_i^0 = \mathbf{x}_i + \mathbf{E}_{\text{pos},i}.
\end{equation*}

where $\mathbf{E}_{\text{pos},i}$ is the positional encoding for the $i$-th position, defined using the classic sinusoidal encoding:
\begin{equation*}
\mathbf{E}_{\text{pos},i}[k] = 
\begin{cases} 
\sin(i / 10000^{k/d}) & \text{if } k \text{ is even}, \\
\cos(i / 10000^{k/d}) & \text{if } k \text{ is odd}.
\end{cases}
\end{equation*}

The embeddings $\mathbf{z}_i^0$ are subsequently processed through a series of $L$ transformer layers. Each layer consists of a multi-head attention (MHA) block and a multilayer perceptron (MLP) block. Layer normalization (LN) is applied before each block, and residual connections are applied after each block.

For the $l$-th layer, 
\begin{equation*}
\mathbf{h}_i^{l} = \text{MHA}(\text{LN}(\mathbf{z}_i^{l-1})) + \mathbf{z}_i^{l-1}.
\end{equation*}
\begin{equation*}
\mathbf{z}_i^{l} = \text{MLP}(\text{LN}(\mathbf{h}_i^{l})) + \mathbf{h}_i^{l}.
\end{equation*}

where,
\begin{equation*}
\text{MHA}(\mathbf{x}) = \text{Concat}(\text{head}_1, \text{head}_2, \dots, \text{head}_h) \mathbf{W}_o,
\end{equation*}
where each attention head $\text{head}_j$ is computed as:
\begin{equation*}
\text{head}_j = \text{softmax}\left(\frac{\mathbf{Q} \mathbf{K}^\top}{\sqrt{d_k}}\right)\mathbf{V},
\end{equation*}
and $\mathbf{Q}, \mathbf{K}, \mathbf{V}$ are query, key, and value matrices derived from $\mathbf{z}$.

The MLP block consists of two fully connected layers with a non-linear Gaussian error linear unit (GELU) activation function:
\begin{equation*}
\text{MLP}(\mathbf{x}) = \text{FC}_2(\text{GELU}(\text{FC}_1(\mathbf{x}))).
\end{equation*}

The final output of the transformer, $\mathbf{z}_i^L \in \mathbb{R}^d$, is mapped to a vector of logits $\mathbf{y}_i \in \mathbb{R}^V$ of size equal to the vocabulary using a linear projection:
\begin{equation*}
\mathbf{y}_i = \mathbf{W}_o \mathbf{z}_i^L + \mathbf{b}_o,
\end{equation*}

\subsection{Training Parameters}
The model is trained using the Adam optimizer with $\beta_1 = 0.9$, $\beta_2 = 0.999$, and a weight decay of $10^{-3}$. We employ a linear one-cycle learning rate schedule with a maximum learning rate of $2 \times 10^{-4}$. 

The transformer architecture consists of $16$ attention heads and $12$ blocks, with dimension of $d=512$, giving a total of 41 million parameters. The model is trained for $1000$ epochs with a batch size of $256$ on 7 NVIDIA H100 GPUs, taking approximately 36 hours to complete. During training, $10\%$ of the training data is used for validation. Early stopping is applied if no significant improvement is observed in the validation loss for $10$ consecutive epochs.

All experiments are performed using a 16-bit mixed precision setup to accelerate training and reduce memory usage.

\subsection{Relaxation and Convex Hull Evaluation}

The structures generated by Matra-Genoa were first optimized using M3GNet~\cite{chenUniversalGraphDeep2022a}, with the FIRE algorithm~\cite{bitzekStructuralRelaxationMade2006} as implemented in ASE~\cite{hjorthlarsenAtomicSimulationEnvironment2017}. Although M3GNet is not the most accurate uMLIP, it is highly numerically efficient. The geometries relaxed with M3GNet are generally reasonable, but in rare cases, nonphysical forces were observed, leading to the exclusion of the affected structures.

The relaxed geometries were analyzed to ensure structural connectivity by examining the null eigenvalues of the Laplacian matrix. The connectivity matrix was constructed by assuming two atoms are connected if their separation is less than 1.5 times the sum of their covalent radii. Any geometries found to be disjoint were discarded.

Subsequently, the M3GNet-optimized geometries were refined using the ORBITAL uMLIP~\cite{neumannOrbFastScalable2024}. Structures optimized with ORBITAL closely resemble PBE geometries. While ORBITAL is computationally slower due to its higher number of parameters, the initial M3GNet geometries required only 20-50 optimization steps to converge forces to 0.02~eV/\AA.

The ORBITAL-optimized geometries were then cross-referenced with the extensive ORB dataset (see \cref{sec:data}) to identify and exclude duplicate compounds. This database currently comprises approximately 72 million entries, and therefore represents a more severe novelty check, largely surpassing comparisons to ICSD or MP.

The distance of each structure to the convex hull was estimated using an ALIGNN model, which predicts PBE-calculated convex hull distances based on M3GNet-optimized geometries. Trained on a dataset of 3,796,799 entries, the ALIGNN model achieved a validation mean absolute error (MAE) of 0.0181~eV/atom on a separate dataset of 300,000 entries. This approach significantly improves the accuracy of convex hull distance predictions compared to direct energy evaluations from ORBITAL.

\subsection{DFT}
All geometry optimizations and total energy calculations were performed with the code VASP~\cite{kresseEfficientIterativeSchemes1996}.
All parameters for the calculations were chosen to be compatible with the Materials Project database.
The Brillouin zones were sampled by uniform $\Gamma$-centered k-point grids with a density of 1000 k-points per reciprocal atom. The projector augmented wave parameters of VASP version 5.2 with a cutoff of 520 eV were applied. The calculations were converged to forces smaller than 0.005 eV/\r{A}. As exchange-correlation functional the Perdew–Burke–Ernzerhof functional was used, with the addition of a $U$ for some oxides and fluorides as in the Materials Project workflow.

\subsection{Generation}
After training on the aforementioned MPA-S dataset, we use the model and obtained weights to sample new sequences. We use 24 cores and 7 H100 GPUs, and are able to generate 3 million structures within 48 hours. This corresponds to 1000 structures per minute, and includes the process of hashing to removing duplicates with respect to previous generated structures or compounds in the training set.

\subsection{Data availability}
The MatraGenoa3M dataset, which includes the 3 million generated structures discussed in this work, is available at \href{https://doi.org/10.6084/m9.figshare.28271294.v1}{https://doi.org/10.6084/m9.figshare.28271294.v1}.

\section*{Acknowledgments}
P.-P. D.B. is thankful to Younesse Kaddar, Gopeshh Subbaraj, Alex Hernandez-Garcia, Hai-Chen Wang, Théo Cavignac, Victor Trinquet, Matthew L. Evans and Yoshua Bengio for their inspiring discussions and helpful insights.

\bibliography{bib}

\begin{thebibliography}{37}%
\makeatletter
\providecommand \@ifxundefined [1]{%
 \@ifx{#1\undefined}
}%
\providecommand \@ifnum [1]{%
 \ifnum #1\expandafter \@firstoftwo
 \else \expandafter \@secondoftwo
 \fi
}%
\providecommand \@ifx [1]{%
 \ifx #1\expandafter \@firstoftwo
 \else \expandafter \@secondoftwo
 \fi
}%
\providecommand \natexlab [1]{#1}%
\providecommand \enquote  [1]{``#1''}%
\providecommand \bibnamefont  [1]{#1}%
\providecommand \bibfnamefont [1]{#1}%
\providecommand \citenamefont [1]{#1}%
\providecommand \href@noop [0]{\@secondoftwo}%
\providecommand \href [0]{\begingroup \@sanitize@url \@href}%
\providecommand \@href[1]{\@@startlink{#1}\@@href}%
\providecommand \@@href[1]{\endgroup#1\@@endlink}%
\providecommand \@sanitize@url [0]{\catcode `\\12\catcode `\$12\catcode `\&12\catcode `\#12\catcode `\^12\catcode `\_12\catcode `\%12\relax}%
\providecommand \@@startlink[1]{}%
\providecommand \@@endlink[0]{}%
\providecommand \url  [0]{\begingroup\@sanitize@url \@url }%
\providecommand \@url [1]{\endgroup\@href {#1}{\urlprefix }}%
\providecommand \urlprefix  [0]{URL }%
\providecommand \Eprint [0]{\href }%
\providecommand \doibase [0]{https://doi.org/}%
\providecommand \selectlanguage [0]{\@gobble}%
\providecommand \bibinfo  [0]{\@secondoftwo}%
\providecommand \bibfield  [0]{\@secondoftwo}%
\providecommand \translation [1]{[#1]}%
\providecommand \BibitemOpen [0]{}%
\providecommand \bibitemStop [0]{}%
\providecommand \bibitemNoStop [0]{.\EOS\space}%
\providecommand \EOS [0]{\spacefactor3000\relax}%
\providecommand \BibitemShut  [1]{\csname bibitem#1\endcsname}%
\let\auto@bib@innerbib\@empty
\bibitem [{\citenamefont {Hohenberg}\ and\ \citenamefont {Kohn}(1964)}]{hohenbergInhomogeneousElectronGas1964a}%
  \BibitemOpen
  \bibfield  {author} {\bibinfo {author} {\bibfnamefont {P.}~\bibnamefont {Hohenberg}}\ and\ \bibinfo {author} {\bibfnamefont {W.}~\bibnamefont {Kohn}},\ }\bibfield  {title} {\bibinfo {title} {Inhomogeneous {{Electron Gas}}},\ }\href {https://doi.org/10.1103/PhysRev.136.B864} {\bibfield  {journal} {\bibinfo  {journal} {Physical Review}\ }\textbf {\bibinfo {volume} {136}},\ \bibinfo {pages} {B864} (\bibinfo {year} {1964})}\BibitemShut {NoStop}%
\bibitem [{\citenamefont {Kohn}\ and\ \citenamefont {Sham}(1965)}]{kohnSelfConsistentEquationsIncluding1965}%
  \BibitemOpen
  \bibfield  {author} {\bibinfo {author} {\bibfnamefont {W.}~\bibnamefont {Kohn}}\ and\ \bibinfo {author} {\bibfnamefont {L.~J.}\ \bibnamefont {Sham}},\ }\bibfield  {title} {\bibinfo {title} {Self-{{Consistent Equations Including Exchange}} and {{Correlation Effects}}},\ }\href {https://doi.org/10.1103/PhysRev.140.A1133} {\bibfield  {journal} {\bibinfo  {journal} {Physical Review}\ }\textbf {\bibinfo {volume} {140}},\ \bibinfo {pages} {A1133} (\bibinfo {year} {1965})}\BibitemShut {NoStop}%
\bibitem [{\citenamefont {Curtarolo}\ \emph {et~al.}(2013)\citenamefont {Curtarolo}, \citenamefont {Hart}, \citenamefont {Nardelli}, \citenamefont {Mingo}, \citenamefont {Sanvito},\ and\ \citenamefont {Levy}}]{curtaroloHighthroughputHighwayComputational2013a}%
  \BibitemOpen
  \bibfield  {author} {\bibinfo {author} {\bibfnamefont {S.}~\bibnamefont {Curtarolo}}, \bibinfo {author} {\bibfnamefont {G.~L.~W.}\ \bibnamefont {Hart}}, \bibinfo {author} {\bibfnamefont {M.~B.}\ \bibnamefont {Nardelli}}, \bibinfo {author} {\bibfnamefont {N.}~\bibnamefont {Mingo}}, \bibinfo {author} {\bibfnamefont {S.}~\bibnamefont {Sanvito}},\ and\ \bibinfo {author} {\bibfnamefont {O.}~\bibnamefont {Levy}},\ }\bibfield  {title} {\bibinfo {title} {The high-throughput highway to computational materials design},\ }\href {https://doi.org/10.1038/nmat3568} {\bibfield  {journal} {\bibinfo  {journal} {Nature Materials}\ }\textbf {\bibinfo {volume} {12}},\ \bibinfo {pages} {191} (\bibinfo {year} {2013})}\BibitemShut {NoStop}%
\bibitem [{\citenamefont {Pickard}\ and\ \citenamefont {Needs}(2011)}]{pickardInitioRandomStructure2011}%
  \BibitemOpen
  \bibfield  {author} {\bibinfo {author} {\bibfnamefont {C.~J.}\ \bibnamefont {Pickard}}\ and\ \bibinfo {author} {\bibfnamefont {R.~J.}\ \bibnamefont {Needs}},\ }\bibfield  {title} {\bibinfo {title} {{\emph{Ab Initio}} random structure searching},\ }\href {https://doi.org/10.1088/0953-8984/23/5/053201} {\bibfield  {journal} {\bibinfo  {journal} {Journal of Physics: Condensed Matter}\ }\textbf {\bibinfo {volume} {23}},\ \bibinfo {pages} {053201} (\bibinfo {year} {2011})}\BibitemShut {NoStop}%
\bibitem [{\citenamefont {Oganov}\ and\ \citenamefont {Glass}(2006)}]{oganovCrystalStructurePrediction2006}%
  \BibitemOpen
  \bibfield  {author} {\bibinfo {author} {\bibfnamefont {A.~R.}\ \bibnamefont {Oganov}}\ and\ \bibinfo {author} {\bibfnamefont {C.~W.}\ \bibnamefont {Glass}},\ }\bibfield  {title} {\bibinfo {title} {Crystal structure prediction using ab initio evolutionary techniques: {{Principles}} and applications},\ }\href {https://doi.org/10.1063/1.2210932} {\bibfield  {journal} {\bibinfo  {journal} {The Journal of Chemical Physics}\ }\textbf {\bibinfo {volume} {124}},\ \bibinfo {pages} {244704} (\bibinfo {year} {2006})}\BibitemShut {NoStop}%
\bibitem [{\citenamefont {Wang}\ \emph {et~al.}(2010)\citenamefont {Wang}, \citenamefont {Lv}, \citenamefont {Zhu},\ and\ \citenamefont {Ma}}]{wangCrystalStructurePrediction2010}%
  \BibitemOpen
  \bibfield  {author} {\bibinfo {author} {\bibfnamefont {Y.}~\bibnamefont {Wang}}, \bibinfo {author} {\bibfnamefont {J.}~\bibnamefont {Lv}}, \bibinfo {author} {\bibfnamefont {L.}~\bibnamefont {Zhu}},\ and\ \bibinfo {author} {\bibfnamefont {Y.}~\bibnamefont {Ma}},\ }\bibfield  {title} {\bibinfo {title} {Crystal structure prediction via particle-swarm optimization},\ }\href {https://doi.org/10.1103/PhysRevB.82.094116} {\bibfield  {journal} {\bibinfo  {journal} {Physical Review B}\ }\textbf {\bibinfo {volume} {82}},\ \bibinfo {pages} {094116} (\bibinfo {year} {2010})}\BibitemShut {NoStop}%
\bibitem [{\citenamefont {Hautier}\ \emph {et~al.}(2011)\citenamefont {Hautier}, \citenamefont {Fischer}, \citenamefont {Ehrlacher}, \citenamefont {Jain},\ and\ \citenamefont {Ceder}}]{hautierDataMinedIonic2011b}%
  \BibitemOpen
  \bibfield  {author} {\bibinfo {author} {\bibfnamefont {G.}~\bibnamefont {Hautier}}, \bibinfo {author} {\bibfnamefont {C.}~\bibnamefont {Fischer}}, \bibinfo {author} {\bibfnamefont {V.}~\bibnamefont {Ehrlacher}}, \bibinfo {author} {\bibfnamefont {A.}~\bibnamefont {Jain}},\ and\ \bibinfo {author} {\bibfnamefont {G.}~\bibnamefont {Ceder}},\ }\bibfield  {title} {\bibinfo {title} {Data {{Mined Ionic Substitutions}} for the {{Discovery}} of {{New Compounds}}},\ }\href {https://doi.org/10.1021/ic102031h} {\bibfield  {journal} {\bibinfo  {journal} {Inorganic Chemistry}\ }\textbf {\bibinfo {volume} {50}},\ \bibinfo {pages} {656} (\bibinfo {year} {2011})}\BibitemShut {NoStop}%
\bibitem [{\citenamefont {Hoffmann}\ \emph {et~al.}(2019)\citenamefont {Hoffmann}, \citenamefont {Maestrati}, \citenamefont {Sawada}, \citenamefont {Tang}, \citenamefont {Sellier},\ and\ \citenamefont {Bengio}}]{hoffmannDataDrivenApproachEncoding2019a}%
  \BibitemOpen
  \bibfield  {author} {\bibinfo {author} {\bibfnamefont {J.}~\bibnamefont {Hoffmann}}, \bibinfo {author} {\bibfnamefont {L.}~\bibnamefont {Maestrati}}, \bibinfo {author} {\bibfnamefont {Y.}~\bibnamefont {Sawada}}, \bibinfo {author} {\bibfnamefont {J.}~\bibnamefont {Tang}}, \bibinfo {author} {\bibfnamefont {J.~M.}\ \bibnamefont {Sellier}},\ and\ \bibinfo {author} {\bibfnamefont {Y.}~\bibnamefont {Bengio}},\ }\href {https://doi.org/10.48550/ARXIV.1909.00949} {\bibinfo {title} {Data-{{Driven Approach}} to {{Encoding}} and {{Decoding}} 3-{{D Crystal Structures}}}} (\bibinfo {year} {2019})\BibitemShut {NoStop}%
\bibitem [{\citenamefont {Noh}\ \emph {et~al.}(2019)\citenamefont {Noh}, \citenamefont {Kim}, \citenamefont {Stein}, \citenamefont {{Sanchez-Lengeling}}, \citenamefont {Gregoire}, \citenamefont {{Aspuru-Guzik}},\ and\ \citenamefont {Jung}}]{nohInverseDesignSolidState2019a}%
  \BibitemOpen
  \bibfield  {author} {\bibinfo {author} {\bibfnamefont {J.}~\bibnamefont {Noh}}, \bibinfo {author} {\bibfnamefont {J.}~\bibnamefont {Kim}}, \bibinfo {author} {\bibfnamefont {H.~S.}\ \bibnamefont {Stein}}, \bibinfo {author} {\bibfnamefont {B.}~\bibnamefont {{Sanchez-Lengeling}}}, \bibinfo {author} {\bibfnamefont {J.~M.}\ \bibnamefont {Gregoire}}, \bibinfo {author} {\bibfnamefont {A.}~\bibnamefont {{Aspuru-Guzik}}},\ and\ \bibinfo {author} {\bibfnamefont {Y.}~\bibnamefont {Jung}},\ }\bibfield  {title} {\bibinfo {title} {Inverse {{Design}} of {{Solid-State Materials}} via a {{Continuous Representation}}},\ }\href {https://doi.org/10.1016/j.matt.2019.08.017} {\bibfield  {journal} {\bibinfo  {journal} {Matter}\ }\textbf {\bibinfo {volume} {1}},\ \bibinfo {pages} {1370} (\bibinfo {year} {2019})}\BibitemShut {NoStop}%
\bibitem [{\citenamefont {Court}\ \emph {et~al.}(2020)\citenamefont {Court}, \citenamefont {Yildirim}, \citenamefont {Jain},\ and\ \citenamefont {Cole}}]{court3DInorganicCrystal2020b}%
  \BibitemOpen
  \bibfield  {author} {\bibinfo {author} {\bibfnamefont {C.~J.}\ \bibnamefont {Court}}, \bibinfo {author} {\bibfnamefont {B.}~\bibnamefont {Yildirim}}, \bibinfo {author} {\bibfnamefont {A.}~\bibnamefont {Jain}},\ and\ \bibinfo {author} {\bibfnamefont {J.~M.}\ \bibnamefont {Cole}},\ }\bibfield  {title} {\bibinfo {title} {3-{{D Inorganic Crystal Structure Generation}} and {{Property Prediction}} via {{Representation Learning}}},\ }\href {https://doi.org/10.1021/acs.jcim.0c00464} {\bibfield  {journal} {\bibinfo  {journal} {Journal of Chemical Information and Modeling}\ }\textbf {\bibinfo {volume} {60}},\ \bibinfo {pages} {4518} (\bibinfo {year} {2020})}\BibitemShut {NoStop}%
\bibitem [{\citenamefont {Long}\ \emph {et~al.}(2021)\citenamefont {Long}, \citenamefont {Fortunato}, \citenamefont {Opahle}, \citenamefont {Zhang}, \citenamefont {Samathrakis}, \citenamefont {Shen}, \citenamefont {Gutfleisch},\ and\ \citenamefont {Zhang}}]{longConstrainedCrystalsDeep2021b}%
  \BibitemOpen
  \bibfield  {author} {\bibinfo {author} {\bibfnamefont {T.}~\bibnamefont {Long}}, \bibinfo {author} {\bibfnamefont {N.~M.}\ \bibnamefont {Fortunato}}, \bibinfo {author} {\bibfnamefont {I.}~\bibnamefont {Opahle}}, \bibinfo {author} {\bibfnamefont {Y.}~\bibnamefont {Zhang}}, \bibinfo {author} {\bibfnamefont {I.}~\bibnamefont {Samathrakis}}, \bibinfo {author} {\bibfnamefont {C.}~\bibnamefont {Shen}}, \bibinfo {author} {\bibfnamefont {O.}~\bibnamefont {Gutfleisch}},\ and\ \bibinfo {author} {\bibfnamefont {H.}~\bibnamefont {Zhang}},\ }\bibfield  {title} {\bibinfo {title} {Constrained crystals deep convolutional generative adversarial network for the inverse design of crystal structures},\ }\href {https://doi.org/10.1038/s41524-021-00526-4} {\bibfield  {journal} {\bibinfo  {journal} {npj Computational Materials}\ }\textbf {\bibinfo {volume} {7}},\ \bibinfo {pages} {66} (\bibinfo {year} {2021})}\BibitemShut {NoStop}%
\bibitem [{\citenamefont {Nouira}\ \emph {et~al.}(2019)\citenamefont {Nouira}, \citenamefont {Sokolovska},\ and\ \citenamefont {Crivello}}]{nouiraCrystalGANLearningDiscover2019a}%
  \BibitemOpen
  \bibfield  {author} {\bibinfo {author} {\bibfnamefont {A.}~\bibnamefont {Nouira}}, \bibinfo {author} {\bibfnamefont {N.}~\bibnamefont {Sokolovska}},\ and\ \bibinfo {author} {\bibfnamefont {J.-C.}\ \bibnamefont {Crivello}},\ }\href {https://doi.org/10.48550/arXiv.1810.11203} {\bibinfo {title} {{{CrystalGAN}}: {{Learning}} to {{Discover Crystallographic Structures}} with {{Generative Adversarial Networks}}}} (\bibinfo {year} {2019}),\ \Eprint {https://arxiv.org/abs/1810.11203} {arXiv:1810.11203 [cs]} \BibitemShut {NoStop}%
\bibitem [{\citenamefont {Zhao}\ \emph {et~al.}(2021)\citenamefont {Zhao}, \citenamefont {Al-Fahdi}, \citenamefont {Hu}, \citenamefont {Siriwardane}, \citenamefont {Song}, \citenamefont {Nasiri},\ and\ \citenamefont {Hu}}]{zhaoHighThroughputDiscoveryNovel2021a}%
  \BibitemOpen
  \bibfield  {author} {\bibinfo {author} {\bibfnamefont {Y.}~\bibnamefont {Zhao}}, \bibinfo {author} {\bibfnamefont {M.}~\bibnamefont {Al-Fahdi}}, \bibinfo {author} {\bibfnamefont {M.}~\bibnamefont {Hu}}, \bibinfo {author} {\bibfnamefont {E.~M.~D.}\ \bibnamefont {Siriwardane}}, \bibinfo {author} {\bibfnamefont {Y.}~\bibnamefont {Song}}, \bibinfo {author} {\bibfnamefont {A.}~\bibnamefont {Nasiri}},\ and\ \bibinfo {author} {\bibfnamefont {J.}~\bibnamefont {Hu}},\ }\bibfield  {title} {\bibinfo {title} {High-{{Throughput Discovery}} of {{Novel Cubic Crystal Materials Using Deep Generative Neural Networks}}},\ }\href {https://doi.org/10.1002/advs.202100566} {\bibfield  {journal} {\bibinfo  {journal} {Advanced Science}\ }\textbf {\bibinfo {volume} {8}},\ \bibinfo {pages} {2100566} (\bibinfo {year} {2021})}\BibitemShut {NoStop}%
\bibitem [{\citenamefont {Ren}\ \emph {et~al.}(2022)\citenamefont {Ren}, \citenamefont {Tian}, \citenamefont {Noh}, \citenamefont {Oviedo}, \citenamefont {Xing}, \citenamefont {Li}, \citenamefont {Liang}, \citenamefont {Zhu}, \citenamefont {Aberle}, \citenamefont {Sun}, \citenamefont {Wang}, \citenamefont {Liu}, \citenamefont {Li}, \citenamefont {Jayavelu}, \citenamefont {Hippalgaonkar}, \citenamefont {Jung},\ and\ \citenamefont {Buonassisi}}]{renInvertibleCrystallographicRepresentation2022b}%
  \BibitemOpen
  \bibfield  {author} {\bibinfo {author} {\bibfnamefont {Z.}~\bibnamefont {Ren}}, \bibinfo {author} {\bibfnamefont {S.~I.~P.}\ \bibnamefont {Tian}}, \bibinfo {author} {\bibfnamefont {J.}~\bibnamefont {Noh}}, \bibinfo {author} {\bibfnamefont {F.}~\bibnamefont {Oviedo}}, \bibinfo {author} {\bibfnamefont {G.}~\bibnamefont {Xing}}, \bibinfo {author} {\bibfnamefont {J.}~\bibnamefont {Li}}, \bibinfo {author} {\bibfnamefont {Q.}~\bibnamefont {Liang}}, \bibinfo {author} {\bibfnamefont {R.}~\bibnamefont {Zhu}}, \bibinfo {author} {\bibfnamefont {A.~G.}\ \bibnamefont {Aberle}}, \bibinfo {author} {\bibfnamefont {S.}~\bibnamefont {Sun}}, \bibinfo {author} {\bibfnamefont {X.}~\bibnamefont {Wang}}, \bibinfo {author} {\bibfnamefont {Y.}~\bibnamefont {Liu}}, \bibinfo {author} {\bibfnamefont {Q.}~\bibnamefont {Li}}, \bibinfo {author} {\bibfnamefont {S.}~\bibnamefont {Jayavelu}}, \bibinfo {author} {\bibfnamefont {K.}~\bibnamefont {Hippalgaonkar}}, \bibinfo {author} {\bibfnamefont {Y.}~\bibnamefont {Jung}},\ and\ \bibinfo
  {author} {\bibfnamefont {T.}~\bibnamefont {Buonassisi}},\ }\bibfield  {title} {\bibinfo {title} {An invertible crystallographic representation for general inverse design of inorganic crystals with targeted properties},\ }\href {https://doi.org/10.1016/j.matt.2021.11.032} {\bibfield  {journal} {\bibinfo  {journal} {Matter}\ }\textbf {\bibinfo {volume} {5}},\ \bibinfo {pages} {314} (\bibinfo {year} {2022})}\BibitemShut {NoStop}%
\bibitem [{\citenamefont {Jiao}\ \emph {et~al.}(2024)\citenamefont {Jiao}, \citenamefont {Huang}, \citenamefont {Lin}, \citenamefont {Han}, \citenamefont {Chen}, \citenamefont {Lu},\ and\ \citenamefont {Liu}}]{jiaoCrystalStructurePrediction2024}%
  \BibitemOpen
  \bibfield  {author} {\bibinfo {author} {\bibfnamefont {R.}~\bibnamefont {Jiao}}, \bibinfo {author} {\bibfnamefont {W.}~\bibnamefont {Huang}}, \bibinfo {author} {\bibfnamefont {P.}~\bibnamefont {Lin}}, \bibinfo {author} {\bibfnamefont {J.}~\bibnamefont {Han}}, \bibinfo {author} {\bibfnamefont {P.}~\bibnamefont {Chen}}, \bibinfo {author} {\bibfnamefont {Y.}~\bibnamefont {Lu}},\ and\ \bibinfo {author} {\bibfnamefont {Y.}~\bibnamefont {Liu}},\ }\href {https://doi.org/10.48550/arXiv.2309.04475} {\bibinfo {title} {Crystal {{Structure Prediction}} by {{Joint Equivariant Diffusion}}}} (\bibinfo {year} {2024}),\ \Eprint {https://arxiv.org/abs/2309.04475} {arXiv:2309.04475 [cond-mat]} \BibitemShut {NoStop}%
\bibitem [{\citenamefont {Xie}\ \emph {et~al.}(2022)\citenamefont {Xie}, \citenamefont {Fu}, \citenamefont {Ganea}, \citenamefont {Barzilay},\ and\ \citenamefont {Jaakkola}}]{xieCrystalDiffusionVariational2022a}%
  \BibitemOpen
  \bibfield  {author} {\bibinfo {author} {\bibfnamefont {T.}~\bibnamefont {Xie}}, \bibinfo {author} {\bibfnamefont {X.}~\bibnamefont {Fu}}, \bibinfo {author} {\bibfnamefont {O.-E.}\ \bibnamefont {Ganea}}, \bibinfo {author} {\bibfnamefont {R.}~\bibnamefont {Barzilay}},\ and\ \bibinfo {author} {\bibfnamefont {T.}~\bibnamefont {Jaakkola}},\ }\href {https://doi.org/10.48550/arXiv.2110.06197} {\bibinfo {title} {Crystal {{Diffusion Variational Autoencoder}} for {{Periodic Material Generation}}}} (\bibinfo {year} {2022}),\ \Eprint {https://arxiv.org/abs/2110.06197} {arXiv:2110.06197 [cs]} \BibitemShut {NoStop}%
\bibitem [{\citenamefont {Zeni}\ \emph {et~al.}(2025)\citenamefont {Zeni}, \citenamefont {Pinsler}, \citenamefont {Z{\"u}gner}, \citenamefont {Fowler}, \citenamefont {Horton}, \citenamefont {Fu}, \citenamefont {Wang}, \citenamefont {Shysheya}, \citenamefont {Crabb{\'e}}, \citenamefont {Ueda}, \citenamefont {Sordillo}, \citenamefont {Sun}, \citenamefont {Smith}, \citenamefont {Nguyen}, \citenamefont {Schulz}, \citenamefont {Lewis}, \citenamefont {Huang}, \citenamefont {Lu}, \citenamefont {Zhou}, \citenamefont {Yang}, \citenamefont {Hao}, \citenamefont {Li}, \citenamefont {Yang}, \citenamefont {Li}, \citenamefont {Tomioka},\ and\ \citenamefont {Xie}}]{zeniGenerativeModelInorganic2025}%
  \BibitemOpen
  \bibfield  {author} {\bibinfo {author} {\bibfnamefont {C.}~\bibnamefont {Zeni}}, \bibinfo {author} {\bibfnamefont {R.}~\bibnamefont {Pinsler}}, \bibinfo {author} {\bibfnamefont {D.}~\bibnamefont {Z{\"u}gner}}, \bibinfo {author} {\bibfnamefont {A.}~\bibnamefont {Fowler}}, \bibinfo {author} {\bibfnamefont {M.}~\bibnamefont {Horton}}, \bibinfo {author} {\bibfnamefont {X.}~\bibnamefont {Fu}}, \bibinfo {author} {\bibfnamefont {Z.}~\bibnamefont {Wang}}, \bibinfo {author} {\bibfnamefont {A.}~\bibnamefont {Shysheya}}, \bibinfo {author} {\bibfnamefont {J.}~\bibnamefont {Crabb{\'e}}}, \bibinfo {author} {\bibfnamefont {S.}~\bibnamefont {Ueda}}, \bibinfo {author} {\bibfnamefont {R.}~\bibnamefont {Sordillo}}, \bibinfo {author} {\bibfnamefont {L.}~\bibnamefont {Sun}}, \bibinfo {author} {\bibfnamefont {J.}~\bibnamefont {Smith}}, \bibinfo {author} {\bibfnamefont {B.}~\bibnamefont {Nguyen}}, \bibinfo {author} {\bibfnamefont {H.}~\bibnamefont {Schulz}}, \bibinfo {author} {\bibfnamefont {S.}~\bibnamefont {Lewis}}, \bibinfo
  {author} {\bibfnamefont {C.-W.}\ \bibnamefont {Huang}}, \bibinfo {author} {\bibfnamefont {Z.}~\bibnamefont {Lu}}, \bibinfo {author} {\bibfnamefont {Y.}~\bibnamefont {Zhou}}, \bibinfo {author} {\bibfnamefont {H.}~\bibnamefont {Yang}}, \bibinfo {author} {\bibfnamefont {H.}~\bibnamefont {Hao}}, \bibinfo {author} {\bibfnamefont {J.}~\bibnamefont {Li}}, \bibinfo {author} {\bibfnamefont {C.}~\bibnamefont {Yang}}, \bibinfo {author} {\bibfnamefont {W.}~\bibnamefont {Li}}, \bibinfo {author} {\bibfnamefont {R.}~\bibnamefont {Tomioka}},\ and\ \bibinfo {author} {\bibfnamefont {T.}~\bibnamefont {Xie}},\ }\bibfield  {title} {\bibinfo {title} {A generative model for inorganic materials design},\ }\href {https://doi.org/10.1038/s41586-025-08628-5} {\bibfield  {journal} {\bibinfo  {journal} {Nature}\ ,\ \bibinfo {pages} {1}} (\bibinfo {year} {2025})}\BibitemShut {NoStop}%
\bibitem [{\citenamefont {Antunes}\ \emph {et~al.}(2024)\citenamefont {Antunes}, \citenamefont {Butler},\ and\ \citenamefont {{Grau-Crespo}}}]{antunesCrystalStructureGeneration2024}%
  \BibitemOpen
  \bibfield  {author} {\bibinfo {author} {\bibfnamefont {L.~M.}\ \bibnamefont {Antunes}}, \bibinfo {author} {\bibfnamefont {K.~T.}\ \bibnamefont {Butler}},\ and\ \bibinfo {author} {\bibfnamefont {R.}~\bibnamefont {{Grau-Crespo}}},\ }\bibfield  {title} {\bibinfo {title} {Crystal structure generation with autoregressive large language modeling},\ }\href {https://doi.org/10.1038/s41467-024-54639-7} {\bibfield  {journal} {\bibinfo  {journal} {Nature Communications}\ }\textbf {\bibinfo {volume} {15}},\ \bibinfo {pages} {10570} (\bibinfo {year} {2024})}\BibitemShut {NoStop}%
\bibitem [{\citenamefont {Zhu}\ \emph {et~al.}(2024)\citenamefont {Zhu}, \citenamefont {Nong}, \citenamefont {Yamazaki},\ and\ \citenamefont {Hippalgaonkar}}]{zhuWyCrystWyckoffInorganic2024}%
  \BibitemOpen
  \bibfield  {author} {\bibinfo {author} {\bibfnamefont {R.}~\bibnamefont {Zhu}}, \bibinfo {author} {\bibfnamefont {W.}~\bibnamefont {Nong}}, \bibinfo {author} {\bibfnamefont {S.}~\bibnamefont {Yamazaki}},\ and\ \bibinfo {author} {\bibfnamefont {K.}~\bibnamefont {Hippalgaonkar}},\ }\href {https://doi.org/10.48550/arXiv.2311.17916} {\bibinfo {title} {{{WyCryst}}: {{Wyckoff Inorganic Crystal Generator Framework}}}} (\bibinfo {year} {2024}),\ \Eprint {https://arxiv.org/abs/2311.17916} {arXiv:2311.17916 [cond-mat]} \BibitemShut {NoStop}%
\bibitem [{\citenamefont {Chantler}\ \emph {et~al.}(2024)\citenamefont {Chantler}, \citenamefont {Boscherini},\ and\ \citenamefont {Bunker}}]{chantlerInternationalTablesCrystallography2024}%
  \BibitemOpen
  \bibinfo {editor} {\bibfnamefont {C.~T.}\ \bibnamefont {Chantler}}, \bibinfo {editor} {\bibfnamefont {F.}~\bibnamefont {Boscherini}},\ and\ \bibinfo {editor} {\bibfnamefont {B.}~\bibnamefont {Bunker}},\ eds.,\ \href {https://doi.org/10.1107/97809553602060000116} {\emph {\bibinfo {title} {International {{Tables}} for {{Crystallography}}: {{X-ray}} Absorption Spectroscopy and Related Techniques}}},\ \bibinfo {edition} {1st}\ ed.,\ Vol.~\bibinfo {volume} {I}\ (\bibinfo  {publisher} {International Union of Crystallography},\ \bibinfo {address} {Chester, England},\ \bibinfo {year} {2024})\BibitemShut {NoStop}%
\bibitem [{\citenamefont {Gebauer}\ \emph {et~al.}(2020)\citenamefont {Gebauer}, \citenamefont {Gastegger},\ and\ \citenamefont {Sch{\"u}tt}}]{gebauerSymmetryadaptedGeneration3d2020}%
  \BibitemOpen
  \bibfield  {author} {\bibinfo {author} {\bibfnamefont {N.~W.~A.}\ \bibnamefont {Gebauer}}, \bibinfo {author} {\bibfnamefont {M.}~\bibnamefont {Gastegger}},\ and\ \bibinfo {author} {\bibfnamefont {K.~T.}\ \bibnamefont {Sch{\"u}tt}},\ }\href {https://doi.org/10.48550/arXiv.1906.00957} {\bibinfo {title} {Symmetry-adapted generation of 3d point sets for the targeted discovery of molecules}} (\bibinfo {year} {2020}),\ \Eprint {https://arxiv.org/abs/1906.00957} {arXiv:1906.00957 [stat]} \BibitemShut {NoStop}%
\bibitem [{\citenamefont {Fredericks}\ \emph {et~al.}(2021)\citenamefont {Fredericks}, \citenamefont {Parrish}, \citenamefont {Sayre},\ and\ \citenamefont {Zhu}}]{fredericksPyXtalPythonLibrary2021}%
  \BibitemOpen
  \bibfield  {author} {\bibinfo {author} {\bibfnamefont {S.}~\bibnamefont {Fredericks}}, \bibinfo {author} {\bibfnamefont {K.}~\bibnamefont {Parrish}}, \bibinfo {author} {\bibfnamefont {D.}~\bibnamefont {Sayre}},\ and\ \bibinfo {author} {\bibfnamefont {Q.}~\bibnamefont {Zhu}},\ }\bibfield  {title} {\bibinfo {title} {{{PyXtal}}: {{A Python}} library for crystal structure generation and symmetry analysis},\ }\href {https://doi.org/10.1016/j.cpc.2020.107810} {\bibfield  {journal} {\bibinfo  {journal} {Computer Physics Communications}\ }\textbf {\bibinfo {volume} {261}},\ \bibinfo {pages} {107810} (\bibinfo {year} {2021})}\BibitemShut {NoStop}%
\bibitem [{\citenamefont {Momma}\ and\ \citenamefont {Izumi}(2011)}]{mommaVESTA3Threedimensional2011}%
  \BibitemOpen
  \bibfield  {author} {\bibinfo {author} {\bibfnamefont {K.}~\bibnamefont {Momma}}\ and\ \bibinfo {author} {\bibfnamefont {F.}~\bibnamefont {Izumi}},\ }\bibfield  {title} {\bibinfo {title} {{{VESTA}} 3 for three-dimensional visualization of crystal, volumetric and morphology data},\ }\href {https://doi.org/10.1107/S0021889811038970} {\bibfield  {journal} {\bibinfo  {journal} {Journal of Applied Crystallography}\ }\textbf {\bibinfo {volume} {44}},\ \bibinfo {pages} {1272} (\bibinfo {year} {2011})}\BibitemShut {NoStop}%
\bibitem [{\citenamefont {Bang}\ \emph {et~al.}(2014)\citenamefont {Bang}, \citenamefont {Lee},\ and\ \citenamefont {Ok}}]{bangVariableFrameworkStructures2014}%
  \BibitemOpen
  \bibfield  {author} {\bibinfo {author} {\bibfnamefont {S.-e.}\ \bibnamefont {Bang}}, \bibinfo {author} {\bibfnamefont {D.~W.}\ \bibnamefont {Lee}},\ and\ \bibinfo {author} {\bibfnamefont {K.~M.}\ \bibnamefont {Ok}},\ }\bibfield  {title} {\bibinfo {title} {Variable {{Framework Structures}} and {{Centricities}} in {{Alkali Metal Yttrium Selenites}}, {{AY}}({{SeO3}})2 ({{A}} = {{Na}}, {{K}}, {{Rb}}, and {{Cs}})},\ }\href {https://doi.org/10.1021/ic500548v} {\bibfield  {journal} {\bibinfo  {journal} {Inorganic Chemistry}\ }\textbf {\bibinfo {volume} {53}},\ \bibinfo {pages} {4756} (\bibinfo {year} {2014})}\BibitemShut {NoStop}%
\bibitem [{\citenamefont {Bengio}\ \emph {et~al.}(2021)\citenamefont {Bengio}, \citenamefont {Jain}, \citenamefont {Korablyov}, \citenamefont {Precup},\ and\ \citenamefont {Bengio}}]{bengioFlowNetworkBased2021a}%
  \BibitemOpen
  \bibfield  {author} {\bibinfo {author} {\bibfnamefont {E.}~\bibnamefont {Bengio}}, \bibinfo {author} {\bibfnamefont {M.}~\bibnamefont {Jain}}, \bibinfo {author} {\bibfnamefont {M.}~\bibnamefont {Korablyov}}, \bibinfo {author} {\bibfnamefont {D.}~\bibnamefont {Precup}},\ and\ \bibinfo {author} {\bibfnamefont {Y.}~\bibnamefont {Bengio}},\ }\href {https://doi.org/10.48550/arXiv.2106.04399} {\bibinfo {title} {Flow {{Network}} based {{Generative Models}} for {{Non-Iterative Diverse Candidate Generation}}}} (\bibinfo {year} {2021}),\ \Eprint {https://arxiv.org/abs/2106.04399} {arXiv:2106.04399 [cs]} \BibitemShut {NoStop}%
\bibitem [{\citenamefont {Jain}\ \emph {et~al.}(2013)\citenamefont {Jain}, \citenamefont {Ong}, \citenamefont {Hautier}, \citenamefont {Chen}, \citenamefont {Richards}, \citenamefont {Dacek}, \citenamefont {Cholia}, \citenamefont {Gunter}, \citenamefont {Skinner}, \citenamefont {Ceder},\ and\ \citenamefont {Persson}}]{jainMaterialsProjectMaterials2013}%
  \BibitemOpen
  \bibfield  {author} {\bibinfo {author} {\bibfnamefont {A.}~\bibnamefont {Jain}}, \bibinfo {author} {\bibfnamefont {S.~P.}\ \bibnamefont {Ong}}, \bibinfo {author} {\bibfnamefont {G.}~\bibnamefont {Hautier}}, \bibinfo {author} {\bibfnamefont {W.}~\bibnamefont {Chen}}, \bibinfo {author} {\bibfnamefont {W.~D.}\ \bibnamefont {Richards}}, \bibinfo {author} {\bibfnamefont {S.}~\bibnamefont {Dacek}}, \bibinfo {author} {\bibfnamefont {S.}~\bibnamefont {Cholia}}, \bibinfo {author} {\bibfnamefont {D.}~\bibnamefont {Gunter}}, \bibinfo {author} {\bibfnamefont {D.}~\bibnamefont {Skinner}}, \bibinfo {author} {\bibfnamefont {G.}~\bibnamefont {Ceder}},\ and\ \bibinfo {author} {\bibfnamefont {K.~A.}\ \bibnamefont {Persson}},\ }\bibfield  {title} {\bibinfo {title} {The {{Materials Project}}: {{A}} materials genome approach to accelerating materials innovation},\ }\href {https://doi.org/10.1063/1.4812323} {\bibfield  {journal} {\bibinfo  {journal} {APL Materials}\ }\textbf {\bibinfo {volume} {1}},\ \bibinfo {pages} {011002}
  (\bibinfo {year} {2013})}\BibitemShut {NoStop}%
\bibitem [{Mat()}]{MatbenchDiscovery}%
  \BibitemOpen
  \href@noop {} {\bibinfo {title} {Matbench {{Discovery}}}},\ \bibinfo {howpublished} {https://matbench-discovery.materialsproject.org/}\BibitemShut {NoStop}%
\bibitem [{\citenamefont {Schmidt}\ \emph {et~al.}(2023)\citenamefont {Schmidt}, \citenamefont {Hoffmann}, \citenamefont {Wang}, \citenamefont {Borlido}, \citenamefont {Carri{\c c}o}, \citenamefont {Cerqueira}, \citenamefont {Botti},\ and\ \citenamefont {Marques}}]{schmidtMachineLearningAssistedDeterminationGlobal2023}%
  \BibitemOpen
  \bibfield  {author} {\bibinfo {author} {\bibfnamefont {J.}~\bibnamefont {Schmidt}}, \bibinfo {author} {\bibfnamefont {N.}~\bibnamefont {Hoffmann}}, \bibinfo {author} {\bibfnamefont {H.-C.}\ \bibnamefont {Wang}}, \bibinfo {author} {\bibfnamefont {P.}~\bibnamefont {Borlido}}, \bibinfo {author} {\bibfnamefont {P.~J. M.~A.}\ \bibnamefont {Carri{\c c}o}}, \bibinfo {author} {\bibfnamefont {T.~F.~T.}\ \bibnamefont {Cerqueira}}, \bibinfo {author} {\bibfnamefont {S.}~\bibnamefont {Botti}},\ and\ \bibinfo {author} {\bibfnamefont {M.~A.~L.}\ \bibnamefont {Marques}},\ }\bibfield  {title} {\bibinfo {title} {Machine-{{Learning}}-{{Assisted Determination}} of the {{Global Zero}}-{{Temperature Phase Diagram}} of {{Materials}}},\ }\href {https://doi.org/10.1002/adma.202210788} {\bibfield  {journal} {\bibinfo  {journal} {Advanced Materials}\ }\textbf {\bibinfo {volume} {35}},\ \bibinfo {pages} {2210788} (\bibinfo {year} {2023})}\BibitemShut {NoStop}%
\bibitem [{\citenamefont {Togo}\ \emph {et~al.}(2024)\citenamefont {Togo}, \citenamefont {Shinohara},\ and\ \citenamefont {Tanaka}}]{togoSpglibSoftwareLibrary2024}%
  \BibitemOpen
  \bibfield  {author} {\bibinfo {author} {\bibfnamefont {A.}~\bibnamefont {Togo}}, \bibinfo {author} {\bibfnamefont {K.}~\bibnamefont {Shinohara}},\ and\ \bibinfo {author} {\bibfnamefont {I.}~\bibnamefont {Tanaka}},\ }\bibfield  {title} {\bibinfo {title} {Spglib: A software library for crystal symmetry search},\ }\href {https://doi.org/10.1080/27660400.2024.2384822} {\bibfield  {journal} {\bibinfo  {journal} {Science and Technology of Advanced Materials: Methods}\ }\textbf {\bibinfo {volume} {4}},\ \bibinfo {pages} {2384822} (\bibinfo {year} {2024})}\BibitemShut {NoStop}%
\bibitem [{\citenamefont {Wang}\ \emph {et~al.}(2021)\citenamefont {Wang}, \citenamefont {Botti},\ and\ \citenamefont {Marques}}]{wangPredictingStableCrystalline2021}%
  \BibitemOpen
  \bibfield  {author} {\bibinfo {author} {\bibfnamefont {H.-C.}\ \bibnamefont {Wang}}, \bibinfo {author} {\bibfnamefont {S.}~\bibnamefont {Botti}},\ and\ \bibinfo {author} {\bibfnamefont {M.~A.~L.}\ \bibnamefont {Marques}},\ }\bibfield  {title} {\bibinfo {title} {Predicting stable crystalline compounds using chemical similarity},\ }\href {https://doi.org/10.1038/s41524-020-00481-6} {\bibfield  {journal} {\bibinfo  {journal} {npj Computational Materials}\ }\textbf {\bibinfo {volume} {7}},\ \bibinfo {pages} {12} (\bibinfo {year} {2021})}\BibitemShut {NoStop}%
\bibitem [{\citenamefont {Bergerhoff}\ \emph {et~al.}(1983)\citenamefont {Bergerhoff}, \citenamefont {Hundt}, \citenamefont {Sievers},\ and\ \citenamefont {Brown}}]{bergerhoffInorganicCrystalStructure1983}%
  \BibitemOpen
  \bibfield  {author} {\bibinfo {author} {\bibfnamefont {G.}~\bibnamefont {Bergerhoff}}, \bibinfo {author} {\bibfnamefont {R.}~\bibnamefont {Hundt}}, \bibinfo {author} {\bibfnamefont {R.}~\bibnamefont {Sievers}},\ and\ \bibinfo {author} {\bibfnamefont {I.~D.}\ \bibnamefont {Brown}},\ }\bibfield  {title} {\bibinfo {title} {The inorganic crystal structure data base},\ }\href {https://doi.org/10.1021/ci00038a003} {\bibfield  {journal} {\bibinfo  {journal} {Journal of Chemical Information and Computer Sciences}\ }\textbf {\bibinfo {volume} {23}},\ \bibinfo {pages} {66} (\bibinfo {year} {1983})}\BibitemShut {NoStop}%
\bibitem [{\citenamefont {Neumann}\ \emph {et~al.}(2024)\citenamefont {Neumann}, \citenamefont {Gin}, \citenamefont {Rhodes}, \citenamefont {Bennett}, \citenamefont {Li}, \citenamefont {Choubisa}, \citenamefont {Hussey},\ and\ \citenamefont {Godwin}}]{neumannOrbFastScalable2024}%
  \BibitemOpen
  \bibfield  {author} {\bibinfo {author} {\bibfnamefont {M.}~\bibnamefont {Neumann}}, \bibinfo {author} {\bibfnamefont {J.}~\bibnamefont {Gin}}, \bibinfo {author} {\bibfnamefont {B.}~\bibnamefont {Rhodes}}, \bibinfo {author} {\bibfnamefont {S.}~\bibnamefont {Bennett}}, \bibinfo {author} {\bibfnamefont {Z.}~\bibnamefont {Li}}, \bibinfo {author} {\bibfnamefont {H.}~\bibnamefont {Choubisa}}, \bibinfo {author} {\bibfnamefont {A.}~\bibnamefont {Hussey}},\ and\ \bibinfo {author} {\bibfnamefont {J.}~\bibnamefont {Godwin}},\ }\href {https://doi.org/10.48550/arXiv.2410.22570} {\bibinfo {title} {Orb: {{A Fast}}, {{Scalable Neural Network Potential}}}} (\bibinfo {year} {2024}),\ \Eprint {https://arxiv.org/abs/2410.22570} {arXiv:2410.22570 [cond-mat]} \BibitemShut {NoStop}%
\bibitem [{Ale()}]{Alexandria}%
  \BibitemOpen
  \href@noop {} {\bibinfo {title} {Alexandria}},\ \bibinfo {howpublished} {https://alexandria.icams.rub.de/pbe/}\BibitemShut {NoStop}%
\bibitem [{\citenamefont {Chen}\ and\ \citenamefont {Ong}(2022)}]{chenUniversalGraphDeep2022a}%
  \BibitemOpen
  \bibfield  {author} {\bibinfo {author} {\bibfnamefont {C.}~\bibnamefont {Chen}}\ and\ \bibinfo {author} {\bibfnamefont {S.~P.}\ \bibnamefont {Ong}},\ }\bibfield  {title} {\bibinfo {title} {A universal graph deep learning interatomic potential for the periodic table},\ }\href {https://doi.org/10.1038/s43588-022-00349-3} {\bibfield  {journal} {\bibinfo  {journal} {Nature Computational Science}\ }\textbf {\bibinfo {volume} {2}},\ \bibinfo {pages} {718} (\bibinfo {year} {2022})}\BibitemShut {NoStop}%
\bibitem [{\citenamefont {Bitzek}\ \emph {et~al.}(2006)\citenamefont {Bitzek}, \citenamefont {Koskinen}, \citenamefont {G{\"a}hler}, \citenamefont {Moseler},\ and\ \citenamefont {Gumbsch}}]{bitzekStructuralRelaxationMade2006}%
  \BibitemOpen
  \bibfield  {author} {\bibinfo {author} {\bibfnamefont {E.}~\bibnamefont {Bitzek}}, \bibinfo {author} {\bibfnamefont {P.}~\bibnamefont {Koskinen}}, \bibinfo {author} {\bibfnamefont {F.}~\bibnamefont {G{\"a}hler}}, \bibinfo {author} {\bibfnamefont {M.}~\bibnamefont {Moseler}},\ and\ \bibinfo {author} {\bibfnamefont {P.}~\bibnamefont {Gumbsch}},\ }\bibfield  {title} {\bibinfo {title} {Structural {{Relaxation Made Simple}}},\ }\href {https://doi.org/10.1103/PhysRevLett.97.170201} {\bibfield  {journal} {\bibinfo  {journal} {Physical Review Letters}\ }\textbf {\bibinfo {volume} {97}},\ \bibinfo {pages} {170201} (\bibinfo {year} {2006})}\BibitemShut {NoStop}%
\bibitem [{\citenamefont {Hjorth~Larsen}\ \emph {et~al.}(2017)\citenamefont {Hjorth~Larsen}, \citenamefont {J{\o}rgen~Mortensen}, \citenamefont {Blomqvist}, \citenamefont {Castelli}, \citenamefont {Christensen}, \citenamefont {Du{\l}ak}, \citenamefont {Friis}, \citenamefont {Groves}, \citenamefont {Hammer}, \citenamefont {Hargus}, \citenamefont {Hermes}, \citenamefont {Jennings}, \citenamefont {Bjerre~Jensen}, \citenamefont {Kermode}, \citenamefont {Kitchin}, \citenamefont {Leonhard~Kolsbjerg}, \citenamefont {Kubal}, \citenamefont {Kaasbjerg}, \citenamefont {Lysgaard}, \citenamefont {Bergmann~Maronsson}, \citenamefont {Maxson}, \citenamefont {Olsen}, \citenamefont {Pastewka}, \citenamefont {Peterson}, \citenamefont {Rostgaard}, \citenamefont {Schi{\o}tz}, \citenamefont {Sch{\"u}tt}, \citenamefont {Strange}, \citenamefont {Thygesen}, \citenamefont {Vegge}, \citenamefont {Vilhelmsen}, \citenamefont {Walter}, \citenamefont {Zeng},\ and\ \citenamefont {Jacobsen}}]{hjorthlarsenAtomicSimulationEnvironment2017}%
  \BibitemOpen
  \bibfield  {author} {\bibinfo {author} {\bibfnamefont {A.}~\bibnamefont {Hjorth~Larsen}}, \bibinfo {author} {\bibfnamefont {J.}~\bibnamefont {J{\o}rgen~Mortensen}}, \bibinfo {author} {\bibfnamefont {J.}~\bibnamefont {Blomqvist}}, \bibinfo {author} {\bibfnamefont {I.~E.}\ \bibnamefont {Castelli}}, \bibinfo {author} {\bibfnamefont {R.}~\bibnamefont {Christensen}}, \bibinfo {author} {\bibfnamefont {M.}~\bibnamefont {Du{\l}ak}}, \bibinfo {author} {\bibfnamefont {J.}~\bibnamefont {Friis}}, \bibinfo {author} {\bibfnamefont {M.~N.}\ \bibnamefont {Groves}}, \bibinfo {author} {\bibfnamefont {B.}~\bibnamefont {Hammer}}, \bibinfo {author} {\bibfnamefont {C.}~\bibnamefont {Hargus}}, \bibinfo {author} {\bibfnamefont {E.~D.}\ \bibnamefont {Hermes}}, \bibinfo {author} {\bibfnamefont {P.~C.}\ \bibnamefont {Jennings}}, \bibinfo {author} {\bibfnamefont {P.}~\bibnamefont {Bjerre~Jensen}}, \bibinfo {author} {\bibfnamefont {J.}~\bibnamefont {Kermode}}, \bibinfo {author} {\bibfnamefont {J.~R.}\ \bibnamefont {Kitchin}}, \bibinfo
  {author} {\bibfnamefont {E.}~\bibnamefont {Leonhard~Kolsbjerg}}, \bibinfo {author} {\bibfnamefont {J.}~\bibnamefont {Kubal}}, \bibinfo {author} {\bibfnamefont {K.}~\bibnamefont {Kaasbjerg}}, \bibinfo {author} {\bibfnamefont {S.}~\bibnamefont {Lysgaard}}, \bibinfo {author} {\bibfnamefont {J.}~\bibnamefont {Bergmann~Maronsson}}, \bibinfo {author} {\bibfnamefont {T.}~\bibnamefont {Maxson}}, \bibinfo {author} {\bibfnamefont {T.}~\bibnamefont {Olsen}}, \bibinfo {author} {\bibfnamefont {L.}~\bibnamefont {Pastewka}}, \bibinfo {author} {\bibfnamefont {A.}~\bibnamefont {Peterson}}, \bibinfo {author} {\bibfnamefont {C.}~\bibnamefont {Rostgaard}}, \bibinfo {author} {\bibfnamefont {J.}~\bibnamefont {Schi{\o}tz}}, \bibinfo {author} {\bibfnamefont {O.}~\bibnamefont {Sch{\"u}tt}}, \bibinfo {author} {\bibfnamefont {M.}~\bibnamefont {Strange}}, \bibinfo {author} {\bibfnamefont {K.~S.}\ \bibnamefont {Thygesen}}, \bibinfo {author} {\bibfnamefont {T.}~\bibnamefont {Vegge}}, \bibinfo {author} {\bibfnamefont {L.}~\bibnamefont
  {Vilhelmsen}}, \bibinfo {author} {\bibfnamefont {M.}~\bibnamefont {Walter}}, \bibinfo {author} {\bibfnamefont {Z.}~\bibnamefont {Zeng}},\ and\ \bibinfo {author} {\bibfnamefont {K.~W.}\ \bibnamefont {Jacobsen}},\ }\bibfield  {title} {\bibinfo {title} {The atomic simulation environment---a {{Python}} library for working with atoms},\ }\href {https://doi.org/10.1088/1361-648X/aa680e} {\bibfield  {journal} {\bibinfo  {journal} {Journal of Physics: Condensed Matter}\ }\textbf {\bibinfo {volume} {29}},\ \bibinfo {pages} {273002} (\bibinfo {year} {2017})}\BibitemShut {NoStop}%
\bibitem [{\citenamefont {Kresse}\ and\ \citenamefont {Furthm{\"u}ller}(1996)}]{kresseEfficientIterativeSchemes1996}%
  \BibitemOpen
  \bibfield  {author} {\bibinfo {author} {\bibfnamefont {G.}~\bibnamefont {Kresse}}\ and\ \bibinfo {author} {\bibfnamefont {J.}~\bibnamefont {Furthm{\"u}ller}},\ }\bibfield  {title} {\bibinfo {title} {Efficient iterative schemes for ab initio total-energy calculations using a plane-wave basis set},\ }\href {https://doi.org/10.1103/PhysRevB.54.11169} {\bibfield  {journal} {\bibinfo  {journal} {Physical Review B}\ }\textbf {\bibinfo {volume} {54}},\ \bibinfo {pages} {11169} (\bibinfo {year} {1996})}\BibitemShut {NoStop}%
\end{thebibliography}%

\clearpage

\onecolumngrid 

\section*{Supplemental Information}
\subsection*{Training data distribution}
\begin{figure}[h!]
    \centering
    \includegraphics[height=1.1\textwidth]{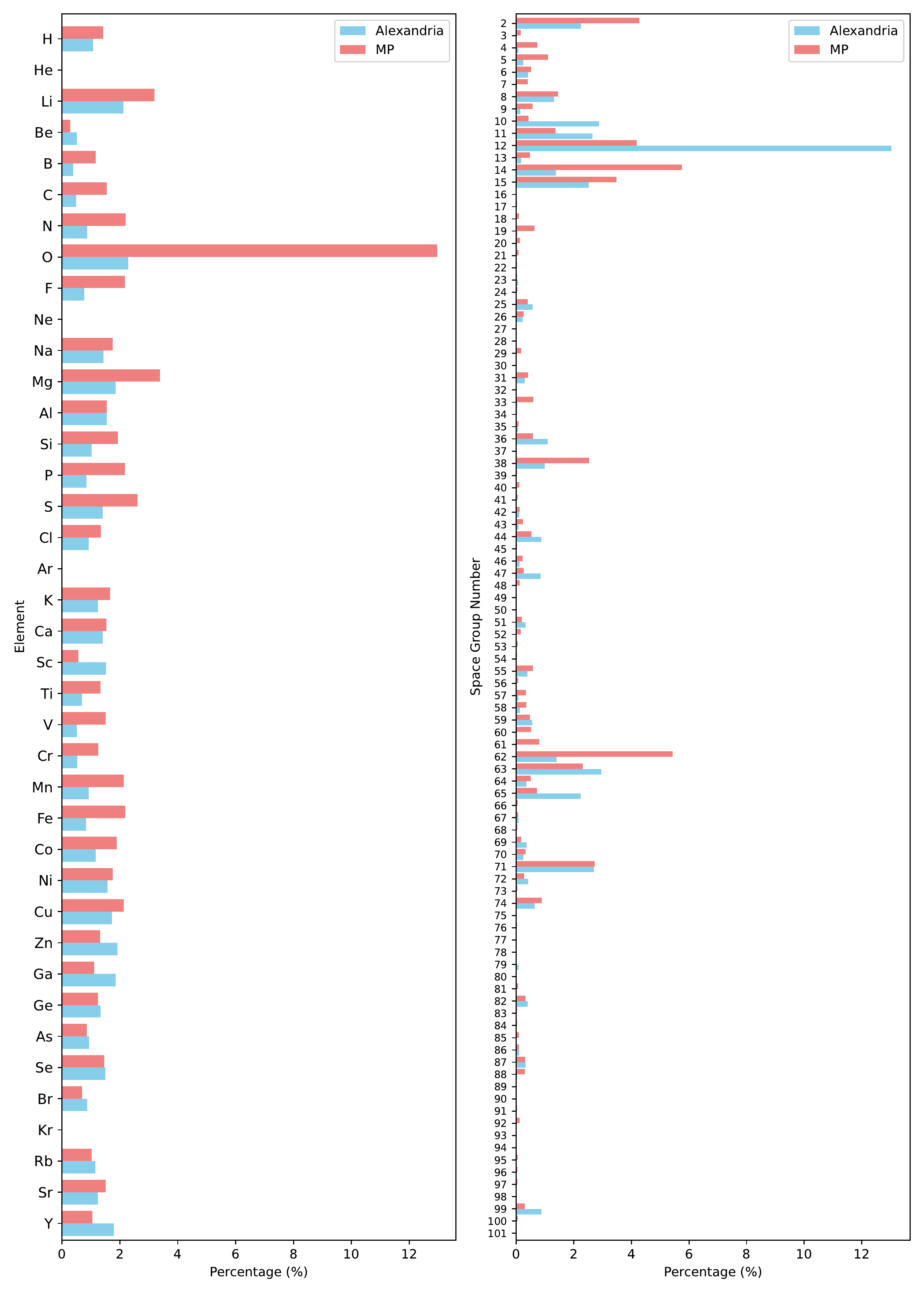}
    \caption{Fractional amount of training samples containing a certain element or having a certain spacegroup. Continued in \cref{fig:train_dist2}.}
    \label{fig:train_dist1}
\end{figure}

\begin{figure}[h!]
    \centering
    \includegraphics[height=1.2\textwidth]{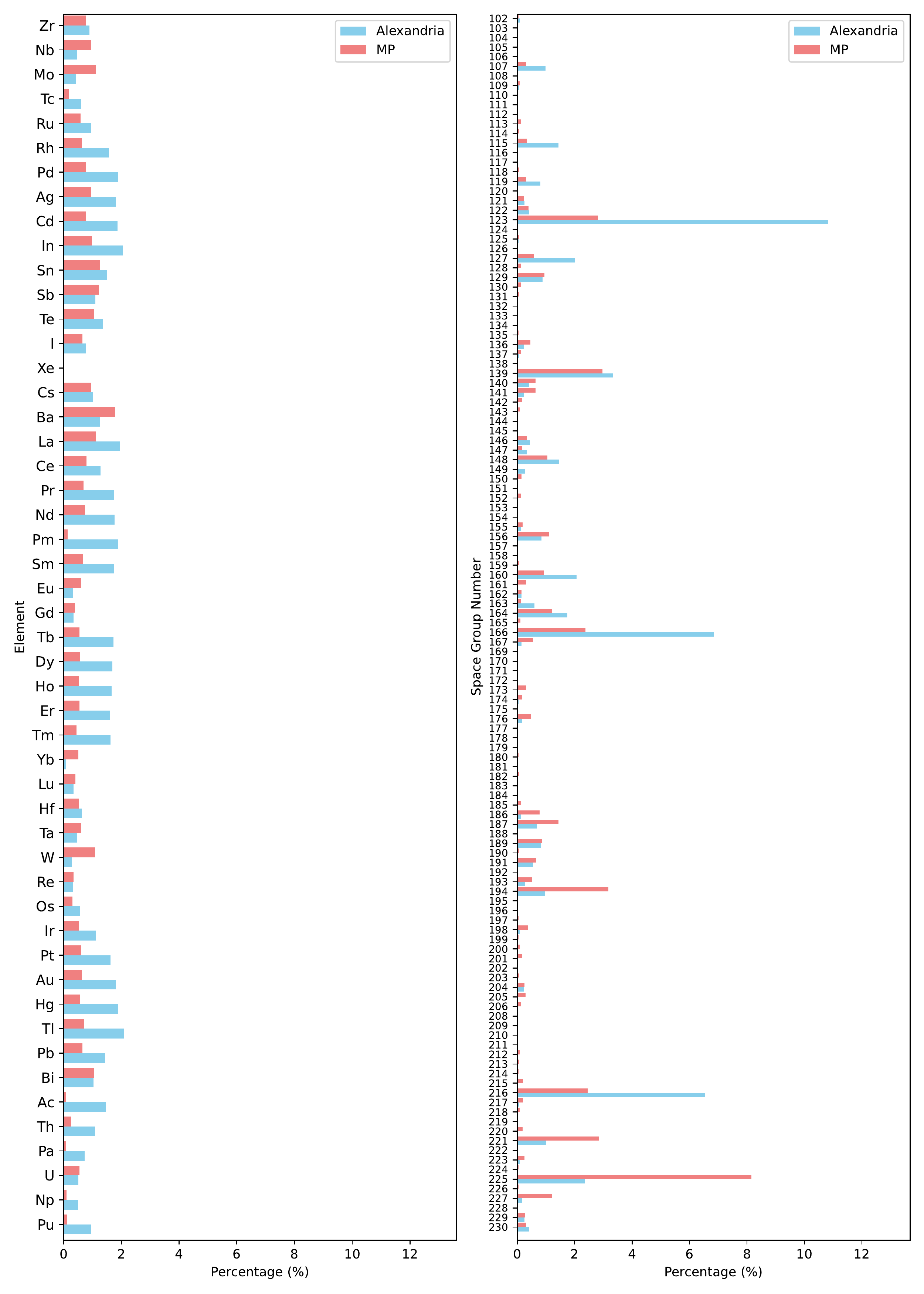}
    \caption{Fractional amount of training samples containing a certain element or having a certain spacegroup. First part in \cref{fig:train_dist1}.}
    \label{fig:train_dist2}
\end{figure}

\subsection*{Learned elemental and spacegroup embeddings}
\begin{figure}[h!]
    \centering
    \includegraphics[width=\textwidth]{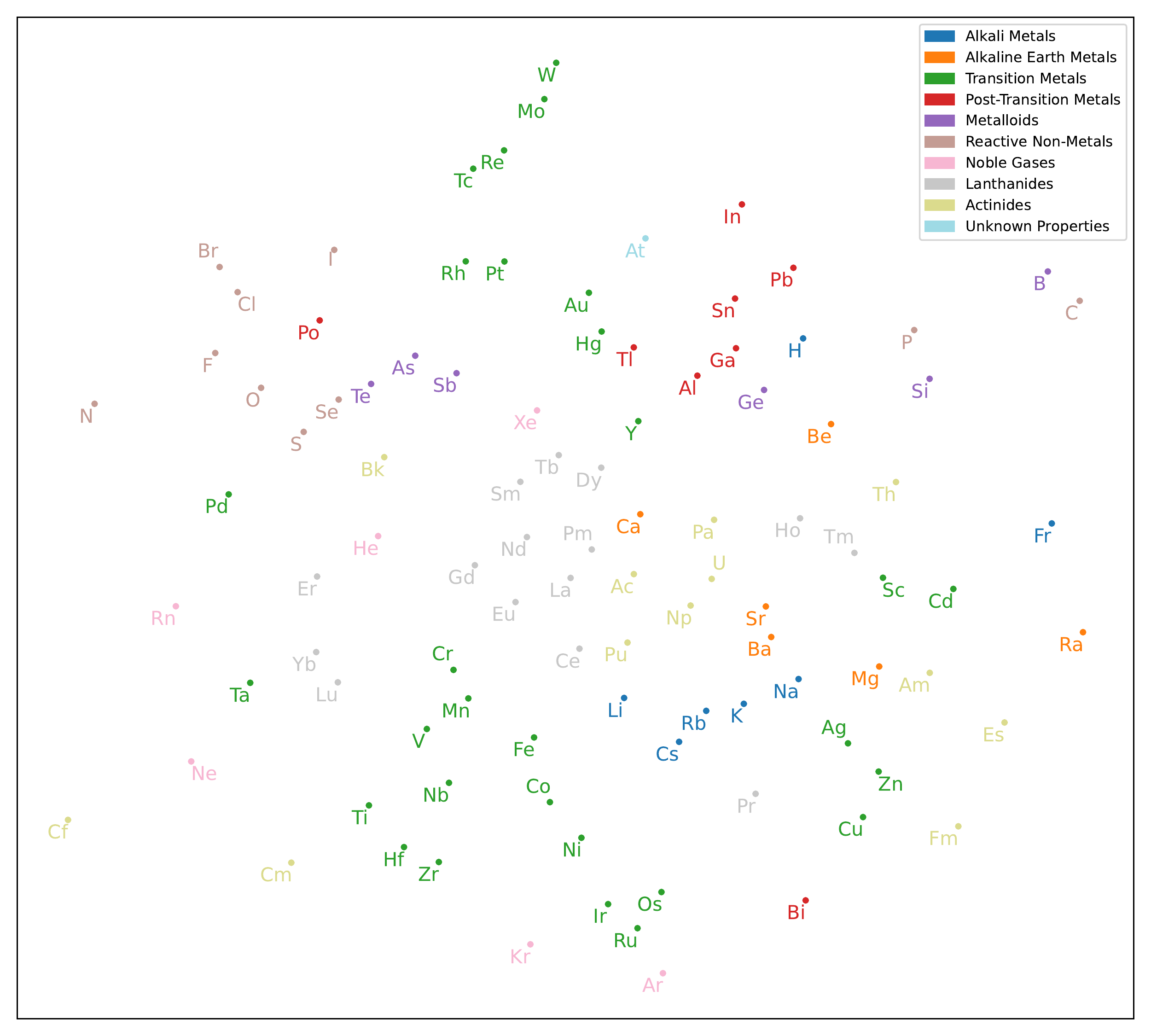}
    \caption{t-SNE dimensionality reduction of the learned embeddings in Matra-Genoa-MPAS, for the different chemical elements, colored by their category. Arbitrary units.}
    \label{fig:elem_embedding}
\end{figure}

\begin{figure}[h!]
    \centering
    \includegraphics[width=\textwidth]{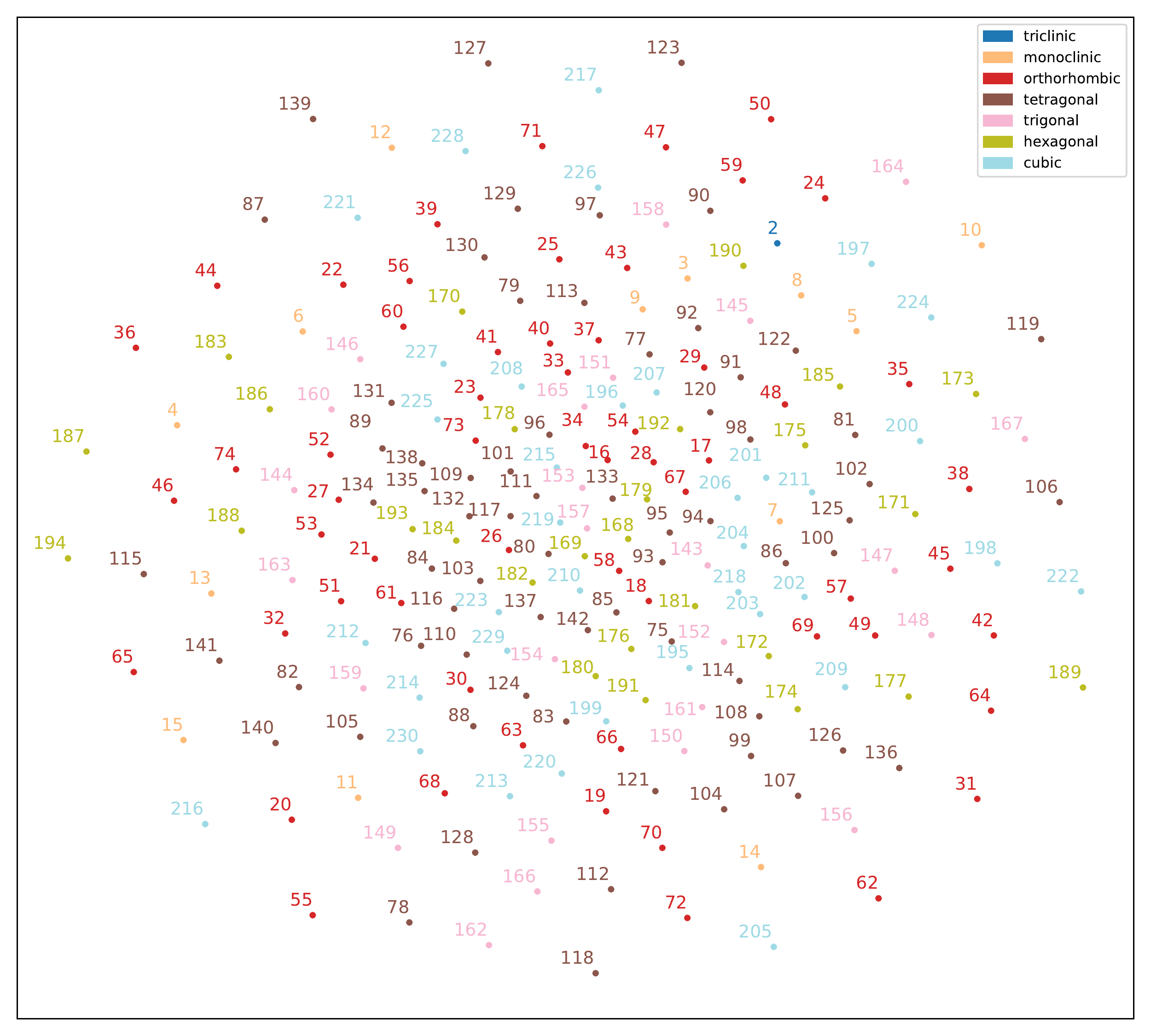}
    \caption{t-SNE dimensionality reduction of the learned embeddings in Matra-Genoa-MPAS, for the different spacegroups, colored by their crystal system. Arbitrary units.}
    \label{fig:spg_embedding}
\end{figure}

\end{document}